\documentclass[acmsmall]{acmart}

\AtBeginDocument{%
  }

\usepackage{graphicx}
\usepackage{multirow}
\usepackage{footnote, url}
\usepackage[flushleft]{threeparttable}
\usepackage{balance}
\usepackage{booktabs}
\usepackage{enumitem}
\usepackage{listings}
\usepackage{subcaption}

\usepackage{amsmath}

\usepackage{soul} %% \st{}

\usepackage{tabularx}           % responsive columns
\newcolumntype{Y}{>{\raggedright\arraybackslash}X}

\usepackage{makecell}

\usepackage{tikz}
\usetikzlibrary{decorations.pathmorphing, shapes}

\newcounter{sarrow}

\newcommand{\accepted}[1]{}

\usepackage{listings}
\usepackage{xcolor}
\usepackage[table]{xcolor}
\usepackage{booktabs}      % for nice rules

\definecolor{codegreen}{rgb}{0,0.6,0}
\definecolor{codegray}{rgb}{0.5,0.5,0.5}
\definecolor{codepurple}{rgb}{0.58,0,0.82}
\definecolor{backcolour}{rgb}{0.95,0.95,0.92}

\lstdefinestyle{mystyle}{
  backgroundcolor=\color{backcolour},   commentstyle=\color{codegreen},
  keywordstyle=\color{magenta},
  numberstyle=\tiny\color{codegray},
  stringstyle=\color{codepurple},
  basicstyle=\ttfamily\footnotesize,
  breakatwhitespace=false,         
  breaklines=true,                 
  captionpos=b,                    
  keepspaces=true,                 
  numbers=left,                    
  numbersep=5pt,                  
  showspaces=false,                
  showstringspaces=false,
  showtabs=false,                  
  tabsize=2
}

\lstdefinestyle{pycode}{
    language=Python,
    basicstyle=\ttfamily\footnotesize,
    keywordstyle=\color{blue}\bfseries,
    commentstyle=\color{gray}\itshape,
    stringstyle=\color{teal},
    showstringspaces=false,
    columns=fullflexible,
    keepspaces=true,
    breaklines=true,
    frame=single
}

\newcommand{\rqone}{\textbf{RQ1:} \textit{How does each type of perturbation in training data affect LLM4Code models’ robustness?}}

\newcommand{\rqtwo}{\textbf{RQ2:} \textit{How does the perturbation ratio affect LLM4Code models’ robustness?}}

\newcommand{\rqthree}{\textbf{RQ3:} \textit{How does the size of perturbed training data affect LLM4Code models’ robustness?
}}

\usepackage[most]{tcolorbox}
\tcbset{colback=gray!5!white, colframe=gray!75!black, fonttitle=\bfseries}

\newtcolorbox{findingbox}[1]{colback=gray!10!white,
  colframe=gray!70!black,
  coltitle=white,
  boxrule=0.8pt,
  arc=2mm,
  left=2mm,
  right=2mm,
  top=1mm,
  bottom=1mm,
  title=Findings (#1)}

\newtcolorbox{Actionable_insight}{colback=gray!10!white,
  colframe=gray!70!black,
  coltitle=white,
  boxrule=0.8pt,
  arc=2mm,
  left=2mm,
  right=2mm,
  top=1mm,
  bottom=1mm,
  title=Actionable Insight}

\usepackage{tcolorbox}
\tcbuselibrary{skins, breakable, theorems}

\newtcbtheorem{answer}{Findings}%
  {enhanced, breakable,
    colback = fbbg, colframe = gray, colbacktitle = fbtitle,
    attach boxed title to top left = {yshift = -2.2mm, xshift = 0mm},
    boxed title style = {rounded corners},
    separator sign none,
    fonttitle = \sffamily\bfseries}{qst}

\usepackage{etoolbox}
\makeatletter
\patchcmd{\@makecaption}
  {\scshape}
  {}
  {}
  {}
\makeatletter
\patchcmd{\@makecaption}
  {\\}
  {.\ }
  {}
  {}
\makeatother

\begin{document}

%%
%% The "title" command has an optional parameter,
%% allowing the author to define a "short title" to be used in page headers.
\title{Improving the Robustness of Large Language Models for Code Tasks via Fine-tuning with Perturbed Data}

%%
%% The "author" command and its associated commands are used to define
%% the authors and their affiliations.
%% Of note is the shared affiliation of the first two authors, and the
%% "authornote" and "authornotemark" commands
%% used to denote shared contribution to the research.
\author{Yang Liu}
%\authornote{Both authors contributed equally to this research.\heng{only one author is marked with *: who is the other author in ``Both authors''}}
\email{yang-2.liu@polymtl.ca}
% \orcid{1234-5678-9012}
\affiliation{%
  \institution{Polytechnique Montréal}
  \city{Montréal}
  \state{Quebec}
  \country{Canada}
}

\author{Armstrong Foundjem}
\affiliation{%
  \institution{Polytechnique Montréal}
  \city{Montréal}
  \state{Quebec}
  \country{Canada}
}

\author{Xingfang Wu}
\affiliation{%
  \institution{Polytechnique Montréal}
  \city{Montréal}
  \state{Quebec}
  \country{Canada}
}

\author{Heng Li}
\affiliation{%
  \institution{Polytechnique Montréal}
  \city{Montréal}
  \state{Quebec}
  \country{Canada}
}

\author{Foutse Khomh}
\affiliation{%
  \institution{Polytechnique Montréal}
  \city{Montréal}
  \state{Quebec}
  \country{Canada}
}

%%
%% By default, the full list of authors will be used in the page
%% headers. Often, this list is too long, and will overlap
%% other information printed in the page headers. This command allows
%% the author to define a more concise list
%% of authors' names for this purpose.
\renewcommand{\shortauthors}{Liu et al.}

%%
%% The abstract is a short summary of the work to be presented in the
%% article.
\begin{abstract}
Context:
In the fast-paced evolution of software development, Large Language Models (LLMs) have become indispensable tools for tasks such as code generation, completion, analysis, and bug fixing. Ensuring the robustness of these models against potential vulnerabilities from handling diverse inputs is critical, as variations in input can lead to incorrect or insecure code outputs.

Objective:
This work aims to improve the robustness of LLMs for coding-related tasks against potential adversarial inputs. Specifically, we investigate how fine-tuning LLMs with perturbed datasets impacts their robustness against input perturbations.

Method:
We systematically evaluated LLM robustness by fine-tuning models using datasets perturbed at character-level, word-level, and sentence-level, comparing results against base models and models fine-tuned on unperturbed datasets.

Results:
Fine-tuning LLMs with perturbed datasets significantly improves model robustness (RD usually drops around 4\% - 6\%), especially for models with relatively weak robustness.
However, this fine-tuning process typically results in a slight performance decrease (pass@1 usually drops around 1\% - 3\%) compared to fine-tuning with unperturbed datasets, although occasional performance improvements are observed.

Conclusion \& Implications:
Fine-tuning LLMs for coding tasks with perturbed data effectively enhances their robustness at the cost of a minor performance reduction, emphasizing the importance of balancing the robustness and performance of LLMs for coding applications.
\end{abstract}

%%
%% The code below is generated by the tool at http://dl.acm.org/ccs.cfm.
%% Please copy and paste the code instead of the example below.
%%


%%
%% Keywords. The author(s) should pick words that accurately describe
%% the work being presented. Separate the keywords with commas.

% \received{20 February 2007}
% \received[revised]{12 March 2009}
% \received[accepted]{5 June 2009}

%%
%% This command processes the author and affiliation and title
%% information and builds the first part of the formatted document.

\maketitle

\section{Introduction}\label{sec:intro}

Large Language Models for Code (LLM4Code)~\cite{nijkamp2022codegen,fried2022incoder} have become indispensable in modern software development~\cite{hou2024large}, supporting tasks such as code generation, completion, analysis, and bug fixing. Beyond efficiency gains, they represent a fundamental shift in development practices~\cite{batouta2016automation}. 
Acting as ``co-developers,'' LLM4Code models assist in solving complex problems and, at times, autonomously generate functional code~\cite{nijkamp2022codegen,fried2022incoder}.
However, their deep integration into the software lifecycle makes them appealing targets for adversaries: carefully crafted inputs or subtle perturbations can trigger incorrect or malicious outputs~\cite{carlini2017towards,metzen2017detecting, jin2020bert,mastropaolo2023robustness}. For example, a minor change in the natural-language problem description, such as replacing ``return the sorted list'' with ``give back the list in order'', can cause some models to ignore edge-case conditions or omit required checks, resulting in syntactically correct but semantically wrong code. Such small linguistic variations, while harmless to a human reader, are sufficient to trigger substantial behavioral shifts in LLM4Code.
% \Foutse{can we give an illustrative example to make this concrete for the reader?}
%However, the increasing reliance on LLM4Code exposes them to adversarial attacks~\cite{jin2020bert,mastropaolo2023robustness}, which can lead to buggy or insecure code. 
Ensuring that these models are resilient is therefore essential for preserving the integrity and security of software systems~\cite{bielik2020adversarial}. 
%Acting as ``co-developers,'' LLM4Code models assist in solving complex problems and, at times, autonomously generate functional code~\cite{nijkamp2022codegen,fried2022incoder}.
%Their deep integration into the software lifecycle makes them appealing targets for adversaries: carefully crafted inputs or subtle perturbations can trigger incorrect or malicious outputs~\cite{carlini2017towards,metzen2017detecting}.

Prior work~\cite{liu2025adversarialattackclassificationrobustness} has shown that the robustness of LLM4Code degrades consistently when prompts are perturbed~\cite{ribeiro-etal-2020-beyond,morris-etal-2020-textattack,wang-etal-2021-textflint}. This fragility presents a significant obstacle to adoption in practical software engineering environments, where inputs often contain variability, ambiguity, or noise. Such behavior can result in syntactically invalid, semantically incorrect, or insecure code, thereby constraining the reliability and maintainability of AI-assisted workflows~\cite{chen2021evaluating,pearce2025asleep}. Moreover, the sensitivity of LLM4Code to small perturbations raises concerns about scalability and its safe integration into large, complex systems. This is because unpredictable or inconsistent model behavior in response to minor input variations can undermine the reliability of automated development pipelines (e.g., Continuous Integration/Deployment). It can also lead to the generation of code that fails subtly in edge cases, increasing debugging costs and operational risks for the entire software system. Addressing robustness~\cite{hendrycks2021measuring} is thus a prerequisite for dependable deployment.

% Building on these challenges \Foutse{To alleviate these issues?}
To alleviate these issues, we investigate how fine-tuning LLMs4Code with perturbed data affects their robustness. Our intuition is that training with controlled noise teaches the model invariances to linguistic variation, a mechanism aligned with adversarial training principles. For instance, Chen et al. (2022) show that structured perturbations encourage more robust and semantically grounded representations, diminishing reliance on superficial patterns~\cite{chen2022adversarial}.
% \yang{Our intuition is that training on controlled noise helps the model learn invariances to linguistic variations, a mechanism supported by adversarial training literature. Chen et al. (2022) demonstrate that training models with structured perturbations promotes the development of more robust, semantically grounded representations, reducing their dependence on brittle surface patterns~\cite{chen2022adversarial}.}
% \Foutse{explain the intuition for pursuing this! Why do we believe that fine-tuning with perturbed data could improve robustness? considering citing some related works in other domains that led you to this intuition...}
% Specifically, we select perturbations at three levels—character, word, and sentence—that significantly impact model robustness~\cite{liu2025adversarialattackclassificationrobustness}. These perturbations are applied individually and in combination to construct 32 perturbed training datasets and 26 perturbed test datasets\Foutse{how are these datasets created?}. We fine-tune base models{\footnote{Base models refer to the original pre-trained model without our fine-tuning.} using both perturbed and unperturbed datasets, and evaluate their robustness under controlled experiments simulating real-world adversarial scenarios~\cite{nijkamp2022codegen,fried2022incoder}.
Specifically, we apply character-, word-, and sentence-level perturbations, which are known to critically impact model robustness~\cite{liu2025adversarialattackclassificationrobustness}, both individually and in combination. This process creates 32 perturbed training and 26 perturbed test datasets; the full construction details are provided in the methodology section. We then fine-tune base models\footnote{Base models refer to the original pre-trained model without our fine-tuning.} on both perturbed and unperturbed datasets and evaluate their robustness in controlled experiments that simulate real-world adversarial scenarios~\cite{nijkamp2022codegen,fried2022incoder}.

\noindent This study addresses the following research questions:
\begin{itemize}
    \item \rqone
    \item \rqtwo
    \item \rqthree
\end{itemize}

Our results show that fine-tuning LLM4Code models with perturbed data improves their robustness, though often at the expense of slightly reduced performance compared to base models and models fine-tuned on unperturbed data\footnote{Our dataset comprises over four programming languages and one HyperText Markup Language, including Python, Java, C++, HTML, and JavaScript. We focus our evaluation on Python. }. 
% \yang{We also experimented on different programming languages besides Python, including Java and C++, and the performance of our models decreases significantly when evaluated on Java and C++ datasets}. For this reason, we use only Python for our experiment.
% \heng{this is confusing since you mentioned "focusing on Python" in the footnote} 
% \armstrong{There is a disconnect here: ``...suggesting language-specific challenges.''} suggesting language-specific challenges.
% \armstrong{Remove this sentence: ``...suggesting language-specific challenges.'' and Add: for this reason, we use only Python for our experiment. Then, fix a similar statement in the conclusion on "different programming languages"}
%%%  \verify{Python, Java, and C++}  \armstrong{5? I can see only three. } 

\noindent \textbf{Key Contributions.}  
This paper makes the following contributions:

\begin{enumerate}
    \item \textbf{Construction of multi-level perturbation datasets.} We design a collection of 32 perturbed training datasets and 26 perturbed test datasets spanning three perturbation levels, providing a systematic dataset for evaluating and improving the robustness of LLM4Code models.
    
    \item \textbf{Robustness-oriented fine-tuning.} 
    % We empirically demonstrate that fine-tuning with perturbed datasets improves robustness across models
    % \heng{from the wording, it is improving the original models, right?}
    % , while quantifying the trade-off between robustness and clean-task performance.%e compared to unperturbed fine-tuning.
    % \st{We show that fine-tuning enhances robustness across base models. Fine-tuning on perturbed datasets yields greater robustness than fine-tuning on unperturbed datasets, with a measurable trade-off in clean-task performance.}
    Our results establish that fine-tuning is essential for building robust models. However, not all fine-tuning is equal. We show that fine-tuning on perturbed data is a more effective strategy for robustness than fine-tuning on clean data, achieving superior resilience at the cost of a slight decrease in performance on clean tasks.

    \item \textbf{Comprehensive robustness analysis.} We examine the effects of perturbation type, ratio, and training dataset\footnote{In this paper, all the training datasets are used for finetuning.} size on LLM4Code robustness, offering quantitative evidence to guide practitioners in balancing performance and resilience.
    
    \item \textbf{Practical deployment recommendations.} We provide actionable insights for deploying LLM4Code in real-world environments where input variability and noise are prevalent.
    
    % \item \textbf{Cross-language robustness evaluation.} We analyze the impact of perturbation-based fine-tuning across multiple programming languages, identifying language-specific challenges and highlighting the difficulty of achieving robustness in multilingual settings.
\end{enumerate}

\noindent \textbf{Paper Organization.} Section~\ref{sec:background} provides background, Section~\ref{sec:related_work} reviews related work, Section~\ref{sec:methodolog} describes our methodology, Section~\ref{sec:results} presents results, Section~\ref{sec:discussion} discusses findings, Section~\ref{sec:implication} summarizes implications, Section~\ref{sec:limi} outlines limitations and future work, Section~\ref{sec:threats} examines threats to validity, and Section~\ref{sec:conclusions} concludes the paper.

\section{Background} \label{sec:background}

Large Language Models for Code (LLM4Code) have achieved remarkable progress in code generation, but their robustness to input variations remains insufficiently addressed. 
While prior work~\cite{liu2025adversarialattackclassificationrobustness} has demonstrated that adversarial or noisy inputs can significantly degrade model performance, systematic defenses against such vulnerabilities remain underdeveloped. 
If these gaps are left unaddressed, LLM4Code systems would be unreliable in practical development environments, where inputs are often noisy, partially structured, or even adversarially crafted, leading to incorrect, insecure, or unstable code outputs.

To motivate our study, we first formalize a \textbf{threat model} that characterizes how input perturbations can be introduced and exploited, either intentionally or unintentionally. 
We then outline \textbf{robustness testing} as a structured methodology to reveal model weaknesses under different perturbation types and levels, which has not been fully established for code generation tasks. 
Finally, we examine \textbf{fine-tuning strategies} as a key mechanism for strengthening model robustness. 
Although fine-tuning has been widely studied in natural language processing, its systematic application to robustness-oriented code generation, generating correct code even when the natural-language specification contains noise, perturbations, or distributional shifts,
% \Foutse{what do you mean by robustness-oriented code generation exactly?}
remains a critical gap that this work aims to fill.

\subsection{Threat Model}
% \st{Our threat model assumes an adversary with limited, black-box access to LLM4Code. The adversary can only interact with the model via crafted input prompts and observe the resulting outputs, without control over the model's internal parameters or architecture.} \yang{\st{A common way of crafting such inputs is by applying perturbations to benign prompts, which has been widely studied as an adversarial strategy in natural language processing and code generation tasks}}%~\cite{wang2022recode}.}
% \armstrong{I make this suggestion: 
Our threat model considers an adversary with \textbf{strictly black-box access} to the target LLM4Code system. The adversary can only interact with the model through its public inference API—submitting crafted input prompts and observing the generated code outputs. This excludes any access to the model's internal parameters, architecture, gradients, or training data. We assume the adversary possesses \textbf{no prior knowledge} of the model's internal architecture or training data. However, they are equipped with domain knowledge about code semantics and common programming errors, enabling the crafting of semantically meaningful perturbations. A practical constraint is that the adversary cannot perform a vast number of queries (e.g., millions) due to typical API rate limits and cost. The primary \textbf{adversarial goal} is to compromise the generated code by inducing one or more of the following:
\begin{itemize}
    \item \textbf{Security Vulnerabilities:} Introducing flaws like SQL injection, buffer overflows, or improper access control.
    \item \textbf{Functional Errors:} Causing the code to produce incorrect outputs or fail at runtime.
    \item \textbf{Specification Deviations:} Leading the model to generate code that does not fulfill the requirements stated in the prompt.
\end{itemize}
A common strategy in this constrained setting involves \textbf{systematically perturbing benign prompts} to exploit model weaknesses. This approach, widely studied in both natural language processing and code generation contexts~\cite{wang2022recode}, allows attackers to search for input variations that trigger harmful outputs without requiring internal model knowledge.

% \heng{A link between crafted input and perturbations is missing, like: A common way of crafting inputs is through creating perturbations to the original benign inputs (provide references).}

We consider three levels of adversarial perturbations~\cite{liu2025adversarialattackclassificationrobustness}:

\begin{itemize}
    \item \textbf{Character-level Perturbations}: Typographical errors and case alterations that mimic common user mistakes or intentional obfuscation.
    \item \textbf{Word-level Perturbations}: Substitutions or insertions of synonyms, which subtly alter semantics while preserving the overall meaning.
    \item \textbf{Sentence-level Perturbations}: Broader transformations, such as back-translation or tense modification, that simulate realistic linguistic variability.
\end{itemize}

The adversary’s primary objective is to exploit model sensitivities to these variations, leading LLM4Code to generate incorrect, insecure, or syntactically flawed code. This threat model reflects practical software development scenarios~\cite{wang2024software} and captures both intentional attacks and unintentional user-induced variations.

\subsection{Robustness Testing}
% Robustness testing is essential for evaluating and improving LLM4Code~\cite{dror2018hitchhiker}. This process involves generating test inputs that capture different types and granularities of adversarial attacks, thereby enabling a systematic assessment of model robustness~\cite{nijkamp2022codegen,fried2022incoder, liu2023your}. \heng{would be good to give a short example of robustness testing} Understanding the nature and impact of perturbations informs the classification of adversarial inputs and guides the development of strategies for improving model reliability\heng{for robustness testing? better use consistent wording to avoid confusion.}. \heng{not quite sure what's the purpose of this sentence: it looks like you want to motivate ``understanding the nature and impact of perturbations'', but this is not the focus of this work}

Robustness testing, which refers to the systematic evaluation of model behavior under controlled input variations,
% \armstrong{What is it, did you define robustness testing above?} 
is essential for evaluating and improving LLM4Code~\cite{dror2018hitchhiker}. This process involves generating test inputs that capture different types and granularities of adversarial attacks, thereby enabling a systematic assessment of model robustness~\cite{nijkamp2022codegen,fried2022incoder, liu2023your}. For example, robustness testing may introduce typographical errors, synonym substitutions, or sentence-level rewriting to examine how well the model responds to perturbed prompts.

\subsection{Fine-tuning}
% Fine-tuning plays a central role in enhancing the robustness of LLM4Code. By retraining models on datasets augmented with perturbed examples, fine-tuning improves their ability to handle variability and ambiguity in user inputs~\cite{sidiropoulos2024improving, tang2023data}.\heng{Is this (fine-tuning with perturbed examples can improve robustness of LLM4Code models) known knowledge or after this work? If it is known knowledge, the novelty of this work is not clear.} This process enables models to generalize across diverse adversarial perturbations, reducing their susceptibility to generating incorrect or insecure code~\cite{bielik2020adversarial,chen2021evaluating}. Beyond defending against known attacks, fine-tuning also enhances stability when facing unseen or noisy inputs, making it a key strategy for deploying LLM4Code in practical development environments.
Fine-tuning, refers to the process of further training a pre-trained model on a task-specific dataset to adapt it to new objectives or domains,
% \armstrong{define it in a simple sentence or cite}
plays a central role in enhancing robustness. In natural language processing, studies have shown that retraining models on datasets augmented with perturbed examples improves resilience to input variability and ambiguity~\cite{sidiropoulos2024improving, tang2023data}. However, the role of fine-tuning with perturbed data in code generation has received far less attention. Understanding its effect in LLM4Code is important, as models must handle diverse adversarial perturbations that can otherwise lead to incorrect or insecure code~\cite{bielik2020adversarial,chen2021evaluating}.

\section{Related Works} \label{sec:related_work}

In this section, we discuss three areas of research on LLM4Code pertinent to our work: the design of pre-trained models, robustness evaluation under adversarial inputs, and fine-tuning strategies for downstream adaptation. We review each stream and highlight how our work advances the state of the art by integrating robustness-oriented fine-tuning with systematic perturbation analysis.

\subsection{Pre-trained Large Language Models for Code}
Pre-trained models have significantly advanced code understanding and generation. Early approaches, such as Code2Vec~\cite{alon2019code2vec} and Code2Seq~\cite{alon2018code2seq}, represent code snippets as embeddings for tasks like method name prediction. While effective, these approaches are not large language models and thus fall outside the current frontier. BERT-style models, including CodeBERT~\cite{feng2020codebert} and GraphCodeBERT~\cite{guo2020graphcodebert}, were subsequently introduced, combining code and natural language training. These models support tasks such as bug fixing and code synthesis but are limited by their short-context capacity. More recently, GPT-based models, such as CodeGen~\cite{nijkamp2022codegen}, StarCoder~\cite{li2023starcoder}, CodeLlama~\cite{roziere2023code}, and DeepSeek-Coder~\cite{guo2024deepseek}, have leveraged massive pre-training corpora and autoregressive modeling, achieving state-of-the-art performance in multi-language code tasks.

Our study focuses on GPT-based models, which represent the dominant family of current LLM4Code. Unlike prior work that primarily evaluates accuracy, we systematically examine their robustness under adversarial perturbations, both before and after fine-tuning on perturbed datasets.

\subsection{Adversarial Attacks and Robustness of LLM4Code}
Robustness evaluation of code models under adversarial attacks has received increasing attention. Wang et al.~\cite{wang2022recode} assessed robustness using perturbations applied directly to code but did not categorize perturbation types or analyze them systematically. Jha et al.~\cite{jha2023codeattack} proposed \textsc{CodeAttack}, a black-box framework that exploits code structure to craft adversarial examples. While effective, their perturbations are structural and overlook natural-language prompt variability. Mastropaolo et al.~\cite{mastropaolo2023robustness} studied GitHub Copilot, showing that minor prompt variations significantly reduce correctness, though their evaluation was restricted to a small set of human-crafted prompts. Improta et al.~\cite{improta2025enhancing} investigated robustness in offensive code generators via data augmentation, emphasizing security-sensitive tasks. However, their approach does not generalize to functional code generation scenarios~\cite{liu2025adversarialattackclassificationrobustness}.

These works collectively highlight the vulnerability of LLM4Code but focus either on narrow perturbation types or test-time evaluation. However, prior works fail to address how to improve the robustness of LLM4Code.
%They did not mention how to improve the robustness.
% \heng{not clear what ``test-time evaluation'' is; do you mean they only focus on evaluating the robustness of LLM4Code models, not on improving the robustness?}
In contrast, our work constructs \textit{multi-level perturbation datasets} (character, word, and sentence)
% \heng{``construction'' is the contribution of your last work, not this one?} 
and studies their effect when used both for evaluation and for robustness-oriented fine-tuning.

\subsection{Fine-tuning LLMs for Code}
Fine-tuning has been widely applied to adapt LLM4Code to specialized tasks. Weyssow et al.~\cite{weyssow2023exploring} investigated parameter-efficient fine-tuning approaches, improving efficiency without explicitly targeting robustness. Li et al.~\cite{li2024fine} explored fine-tuning for secure code generation, focusing on security concerns rather than general robustness. Shi et al.~\cite{shi2023towards} provided a broad empirical comparison of fine-tuning methods, offering efficiency insights but without considering adversarial robustness. He et al.~\cite{he2024instruction} studied instruction tuning for secure code, again limiting their scope to security objectives.

Our work differs from these works by explicitly targeting robustness improvement through fine-tuning on systematically perturbed datasets. We evaluate multiple perturbation levels, perturbation ratios, and dataset sizes, providing empirical insights into how fine-tuning strategies can be adapted to improve robustness in practice. To our knowledge, this is the first study to comprehensively investigate robustness-oriented fine-tuning for LLM4Code.

\subsection{Comparative Positioning}
Taken together, prior research has advanced LLM4Code in three directions: (i) designing stronger pre-trained models, (ii) probing robustness under adversarial conditions, and (iii) exploring efficient fine-tuning strategies. However, no existing work systematically investigates robustness-oriented fine-tuning across multiple perturbation levels and dataset configurations. Our study bridges this gap by combining controlled perturbations with fine-tuning for more robust LLM4Code models, yielding new insights into the trade-offs between robustness and performance.

\section{Methodology} \label{sec:methodolog}

This study proposes a framework for systematically improving the robustness of LLM4Code models when fine-tuned with perturbed datasets. The framework builds on SafeCoder’s instruction-tuning setup~\cite{he2024instruction} and enables controlled experiments across multiple perturbation types, ratios, and dataset sizes. This section describes the framework, datasets, perturbation strategies, and model selection in detail.

\subsection{Framework Overview}
\begin{figure*}[!ht]
    \centering
    \includegraphics[scale=0.5]{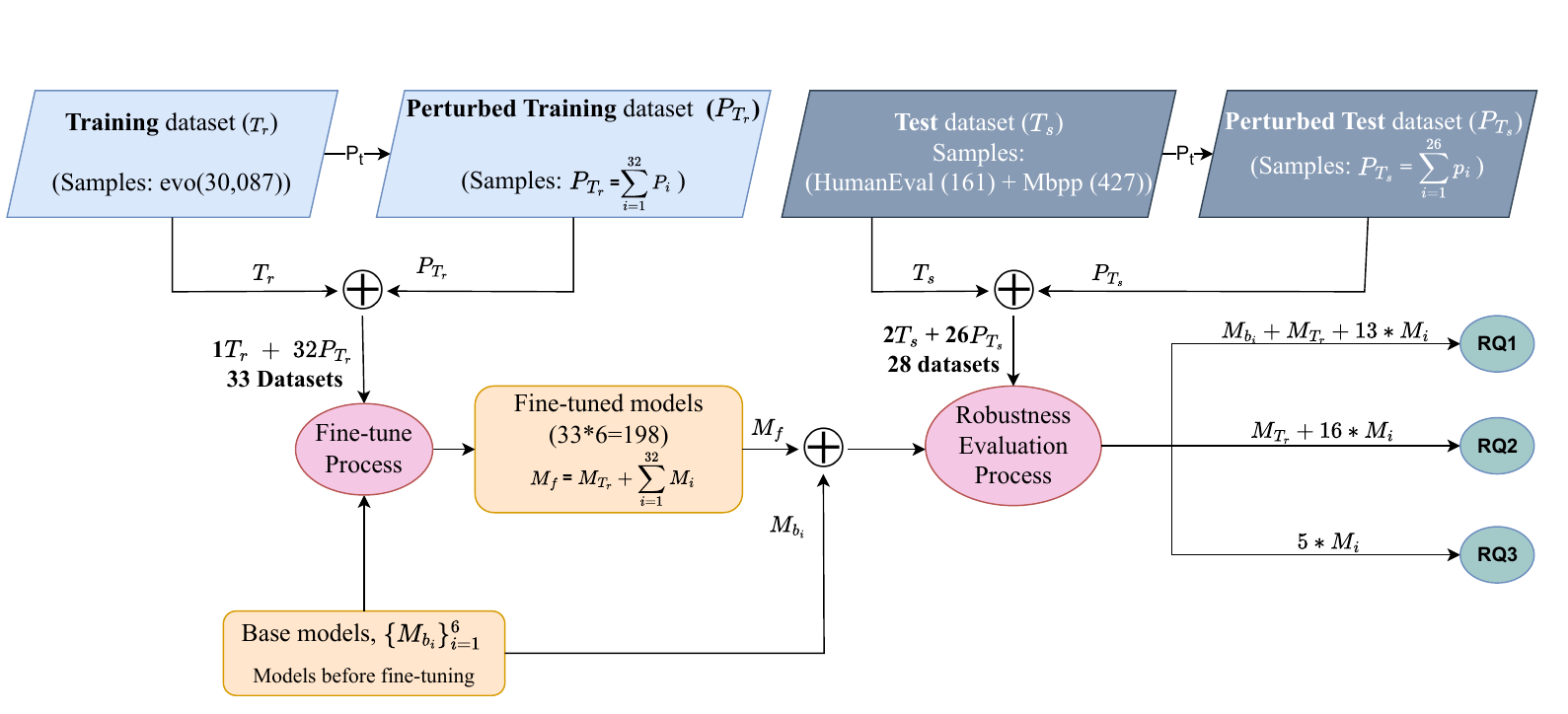}
    \caption[Framework]{\footnotesize Overall framework for robustness-oriented fine-tuning of LLM4Code models and its evaluation
    % \heng{I think the contribution is limited to ``evaluation'', but also the robustness-oriented fine-tuningapproach} 
    the robustness of fine-tuned LLM4Code models. The framework begins with one unperturbed dataset (${T}_r$) and 32 perturbed datasets ($P_{T_r}$). Six baseline models ($M_{b_{1...6}}$) are fine-tuned separately on these datasets, producing 198 fine-tuned models in total. All models (six base
    % \heng{base: baseline}
    and 198 fine-tuned) are then evaluated against $T_s$ and $P_{T_s}$ test sets (HumanEval with 161 tasks and MBPP with 427 tasks), along with their perturbed variants. }
    \label{fig:framework}
\end{figure*}

Figure~\ref{fig:framework} illustrates the overall framework used in this study. We begin by constructing one unperturbed training dataset and 32 perturbed training datasets, generated using character-, word-, and sentence-level perturbation methods. We define six base
% baseline\heng{it is called ``base model'' in Figure 1; ``base model'' looks better}
models,\footnote{Each base model refers to the original, pre-trained LLM4Code prior to fine-tuning. These serve as reference points for comparing fine-tuned models ($M_f$).} denoted as $\{M_{b_i}\}_{i=1}^{6}$. Each baseline model $M_{b_i}$ is fine-tuned on either the unperturbed dataset ($T_r$) or perturbed datasets ($T_r \oplus P_{T_r}$), where $P_{T_r}$ represents the perturbation operator applied to the training set.

\noindent\textbf{Fine-tuning process.}  
For each baseline model, we perform fine-tuning across all available training datasets. This includes one unperturbed dataset and 32 perturbed datasets, resulting in $33$ fine-tuned variants per base model. With six baselines, this yields $6 \times 33 = 198$ fine-tuned models in total. These fine-tuned models form the family of robustness-oriented models ($M_f$).

\noindent\textbf{Evaluation process.}  
% All resulting models are evaluated on both the unperturbed test dataset ($T_s$) and perturbed test datasets ($P_{T_s}$). The evaluation process enables direct comparison of base models, unperturbed fine-tuned models, and perturbation-aware 
% % \Foutse{briefly defines the term "perturbation-aware fine-tuned model"!} 
% fine-tuned models, \yang{which means models fine-tuned on training data that deliberately incorporates controlled perturbations}, under consistent testing conditions.\\  
% \yang{To assess statistical significance, we use paired Wilcoxon signed-rank tests for comparing base, unperturbed fine-tuned, and perturbation-aware fine-tuned models.
% We do not apply Bonferroni or other multiple-comparison corrections, as the hypothesis being tested is identical across all model variants and the comparisons serve as repeated measurements of the same effect rather than independent statistical tests.
% This design follows standard practice in robustness evaluation, where the goal is to assess the consistency of improvement rather than to perform separate hypothesis testing for each model configuration.}\\
All resulting models are evaluated on both the unperturbed test set ($T_s$) and the perturbed test sets ($P_{T_s}$). This enables a direct comparison of base models, models fine-tuned on unperturbed data, and perturbation-aware fine-tuned models (i.e., models fine-tuned on data that deliberately incorporates controlled perturbations) under consistent conditions.

To assess statistical significance, we employ paired Wilcoxon signed-rank tests comparing these model categories. We do not apply multiple-comparison corrections, as we are testing the same core hypothesis—whether fine-tuning improves robustness—across repeated measurements of related model variants. This approach aligns with standard robustness evaluation, which prioritizes measuring the consistency of an effect over performing independent tests for each configuration. 
% \Foutse{Also, how would you do the statistical analysis of your comparisons? do you use some paired-tests? do you correct for p-values given your large number of fine-tuned models variants?}\\
\textbf{RQ-specific configurations.}  
Each research question was designed with a distinct configuration aligned to its objective. The number of datasets used in RQ1 (13), RQ2 (16), and RQ3 (5) (explained in the specific RQs) derives directly from the perturbation design and experimental manipulations, rather than arbitrary choices. These datasets represent overlapping subsets of the same design space, and many fine-tuned models are reused across questions. Consequently, the total number of fine-tuned models is not additive across RQs, but fixed at 33 per base model, yielding 198 models overall. This framework systematically explores three experimental dimensions: (i) the effect of perturbation type (RQ1), (ii) the effect of perturbation ratio (RQ2), and (iii) the effect of training dataset size (RQ3). It provides a controlled setting for robustness evaluation and enables direct comparison among base models, unperturbed fine-tuned models, and perturbation-aware fine-tuned models.

%%\armstrong{How did we define \(M_i\) again? See RQ1-3}
%% $A \oplus B$

% \heng{I think this paragraph should be moved to the previous sub-section (4.1), after the first paragraph, ans use a header like ``Three categories of models''}
\subsubsection{Three categories of models.}To assess whether fine-tuning with perturbed datasets improves model robustness, we evaluate three categories of models: (i) the original base models, (ii) models fine-tuned on the unperturbed training dataset, and (iii) models fine-tuned on perturbed training datasets. This configuration was chosen to establish a clear comparative baseline and to isolate the effect of perturbation-aware fine-tuning. Evaluating the original base models (i) provides a reference point for the inherent robustness of pre-trained LLM4Code. Fine-tuning on the unperturbed dataset (ii) reflects standard adaptation practices and enables us to quantify robustness gains or losses attributable solely to additional exposure to perturbations. Finally, fine-tuning on perturbed datasets (iii) allows us to directly test whether robustness can be systematically improved by training with adversarially varied inputs. Together, these three categories create a controlled experimental setup that balances validity, interpretability, and reproducibility.  
We note, however, that this design does not exhaustively explore hybrid strategies beyond the perturbation ratios later introduced in Table~\ref{tab:classification}. For instance, dynamic curricula or adversarial training schedules could also be considered. Our chosen configuration focuses on clarity and experimental tractability, ensuring that robustness effects are attributable to perturbation-aware fine-tuning rather than to more complex training dynamics.

\subsection{Research Questions (RQs)}\label{sec:rq}
\textbf{\rqone}

% \textbf{Objective:} This research question focuses on exploring the impact of perturbation types on robustness of models, including the impact of each single perturbation and multi-perturbations. We would like to explore the differences in their impacts for perturbations of the same level, and the differences in mixed perturbations in this research question.
\textbf{Objective.}  
This research question aims to assess how different perturbation types influence model robustness. As shown in Table~\ref{tab:classification}, we examine both single perturbations (character-, word-, and sentence-level)\cite{liu2025adversarialattackclassificationrobustness} and multi-perturbations (combinations of types), comparing their relative effects. Specifically, we seek to determine (i) whether perturbations at the same level differ in impact, and (ii) whether training on mixed perturbations yields greater robustness than training on individual types.

\textbf{Approach:} Our fine-tuning strategy, illustrated in Figure~\ref{fig:framework}, is designed to evaluate the impact of input perturbations on model robustness. For each of the six base models (\(M_{b_{1...6}}\)), we create a series of 14 fine-tuned variants.

% \armstrong{ I make this suggstion: \\
This is achieved by fine-tuning each base model on a distinct dataset derived from the original training set (\(T_r\)). These datasets comprise the unperturbed \(T_r\) and 13 perturbed versions (\(P_{T_r}\)). The 13 perturbed datasets are constructed using the methods categorized in Table~\ref{tab:classification}:
\begin{itemize}
    \item \textbf{Character-level (C):} Three datasets generated via typographical perturbations (C1--C3).
    \item \textbf{Word-level (W):} Three datasets generated via semantic-preserving word changes (W1--W3).
    \item \textbf{Sentence-level (S):} Three datasets generated via syntactic rephrasing (S1--S3).
    \item \textbf{Aggregate Datasets:} Four additional datasets are created by combining the above: one for each granularity level (C, W, S) and one final dataset combining all perturbations.
\end{itemize}
Consequently, for each base model, we produce 14 fine-tuned models (\(M_f\))—one for each training dataset—resulting in a total of 15 model instances per base model when including the original.

To evaluate our RQ1, we ensure:

\noindent -- The base model ($M_{b_{1...6}}$) and the unperturbed fine-tuned model ($M_{T_r}$) are tested on all test datasets ($T_s$ and all $P_{T_s}$).\\
-- Each single-perturbation fine-tuned model ($M_f$) is tested on the unperturbed test dataset ($T_s$) and the perturbed test datasets ($P_{T_s}$) corresponding to the same perturbation level. \\
-- Each multi-perturbation fine-tuned model ($M_f$) is tested on the unperturbed test dataset ($T_s$) and on all multi-perturbation test datasets ($P_{T_s}$) that combine perturbations across different levels. 
In this work, single-perturbation settings are not meant to directly mirror deployment conditions. 
Instead, they serve as controlled probes for isolating the effect of each perturbation type on robustness and for building intuition about which forms of input noise are most harmful, before moving to the more realistic multi-perturbation scenario.
This design reflects our main focus on (i) understanding how individual perturbation \emph{types} affect robustness in controlled settings, and (ii) evaluating a more realistic scenario where inputs contain heterogeneous noise rather than a single isolated perturbation. 
We therefore do not further compare multi-perturbation fine-tuned models on single-perturbation test sets, as this cross-setting analysis is beyond the primary scope of our research questions. 

% \Foutse{so we never compare whether fine-tuning on multi-perturbation yields more robustness on single perturbation test sets? In comparison to fine-tuning on single perturbation?}

This setup, consistent with the RQ1 robustness evaluation process in Figure~\ref{fig:framework}, ensures that evaluation is aligned with the type of perturbation applied during fine-tuning while maintaining a common baseline comparison on the unperturbed test dataset.

\begin{table*}[!ht]
\centering
\caption{Nine perturbation methods grouped into three categories; character-level (C1–C3) introduces typographical noise, word-level (W1–W3) alters lexical choices, and sentence-level (S1–S3) applies broader semantic shifts. These methods simulate realistic variations, i.e.\ adversarial manipulations in NL prompts.}
\label{tab:classification}
\resizebox{\linewidth}{!}{%
\begin{tabular}{l|p{9cm}|l|c|c}
\toprule
\textbf{Perturbation Method} & \textbf{Definition} & \textbf{Category} & \textbf{Group (C/W/S)} & \textbf{Group Total} \\
\midrule
ButterFingers (C1)            & Random character insertions/typos. & Character-level & \multirow{3}{*}{C} & \multirow{9}{*}{\makecell{Aggregate dataset composition: \\ 9 individual methods (C1--S3) \\ + 3 category aggregates (C+W+S) \\ + 1 combined dataset (100\%) \\ \textbf{Total: 13 datasets}}} \\
SwapCharacters (C2)           & Character swaps within words.      & Character-level &                      &                              \\
ChangeCharCase (C3)           & Case modifications.                & Character-level &                      &                              \\
\cline{1-4}
SynonymInsertion (W1)         & Insertion of synonyms.             & Word-level      & \multirow{3}{*}{W} &                              \\
SynonymSubstitution (W2)      & Replacement with synonyms.         & Word-level      &                      &                              \\
InflectionalVariation (W3)    & Word inflection changes.           & Word-level      &                      &                              \\
\cline{1-4}
BackTranslation (S1)          & Round-trip translation.            & Sentence-level  & \multirow{3}{*}{S} &                              \\
TenseTransformationPast (S2)  & Conversion to past tense.          & Sentence-level  &                      &                              \\
TenseTransformationFuture (S3)& Conversion to future tense.        & Sentence-level  &                      &                              \\
\bottomrule
\end{tabular}}
\end{table*}

\textbf{\rqtwo}

\textbf{Objective:} This research question focuses on exploring the impact of the proportion of perturbed samples in the training datasets on the robustness of LLM4Code models. 
% We would like to get a great proportion\heng{not clear what ``a great proportion'' means} in this research question.

% \textbf{Approach:} We randomly selected samples in original training dataset and nine perturbed training datasets in three times. The proportion is 0\%, 10\%, 30\%, 50\%, 70\%, 90\% and 100\%. 0\% is the unperturbed dataset, and 100\% is the same dataset in RQ1 with all perturbation methods. For other datasets, we repeat the randomly select process three times in order to reduce the impact of accidental. All the fine-tuned models are evaluated by the unperturbed test dataset and perturbed test dataset including all perturbation methods.
\textbf{Approach:} As illustrated in Figure~\ref{fig:framework}, for each base model ($M_{b_{1...6}}$), we construct mixed training datasets by combining unperturbed samples ($T_r$) and perturbed samples ($P_{T_r}$) at different proportions: 0\%, 10\%, 30\%, 50\%, 70\%, 90\%, and 100\%. 
Here, 0\% corresponds to a fully unperturbed dataset ($T_r$) and 100\% corresponds to the dataset in RQ1 containing only perturbed data ($P_{T_r}$). The perturbation ratios (10\%, 30\%, 50\%, 70\%, 90\%) 
% \armstrong{add: 10\%, 30\%, 50\%, 70\%, 90\%} 
were chosen to balance experimental coverage with computational feasibility. 
% \armstrong{Explain how the 16 subsets were obtained 16: 3 (repeated runs) x 5 (10\%, 30\%, 50\%, 70\%, 90\%) + 1 (100\%)}
We sampled the space more densely at the extremes (0\%, 10\%, 90\%, 100\%) to capture boundary effects, 
where small changes in the proportion of perturbed data may have a disproportionate impact on robustness. 
In the middle range, we used coarser steps (30\%, 50\%, 70\%) to provide representative coverage without exhaustively testing every possible ratio. 
% This design allows us to capture both edge-case sensitivity and overall trends, while keeping the total number of fine-tuned models computationally manageable. Intermediate proportions are generated by random sampling, and to reduce stochastic effects, the sampling process is repeated three times. \yang{In total, this yields 16 fine-tuned models for RQ2: one trained on fully perturbed data (100\%), and 15 models obtained from three repeated runs at five intermediate perturbation ratios (3 (repeated runs) x 5 (10\%, 30\%, 50\%, 70\%, 90\%) + 1 (100\%)).} 
% % \heng{Lack an explanation of why there are 16 models for RQ2}
% Each resulting fine-tuned model ($M_f$) is then evaluated on both the unperturbed test dataset($T_s$) and all perturbed test datasets ($P_{T_s}$), enabling systematic comparison across different perturbation ratios.
This design captures both edge-case sensitivity and broader performance trends within a computationally feasible total of models. We generate intermediate proportions by random sampling and repeat this process three times at each level to mitigate stochastic effects. In total, this yields 16 fine-tuned models for RQ2: one trained on fully perturbed data (100\%), plus 15 models from three repetitions at each of five intermediate perturbation ratios—10\%, 30\%, 50\%, 70\%, and 90\% (3 repetitions × 5 ratios + 1 full-perturbation model).

Each resulting fine-tuned model ($M_f$) is evaluated on both the unperturbed test set ($T_s$) and all perturbed test sets ($P_{T_s}$), allowing a systematic comparison across perturbation levels.

\textbf{\rqthree}

\textbf{Objective:} This research question focuses on exploring the samples size in training dataset on robustness of models. We would like to configure the range to avoid overfitting and underfitting in this research question.

% \textbf{Approach:} We randomly choose one-quarter, one-half, two times, and three times the number of samples in perturbed training datasets with nine perturbation methods. All the fine-tuned models are evaluated by the unperturbed test dataset and perturbed test dataset including all perturbation methods.
\textbf{Approach:} As illustrated in Figure~\ref{fig:framework}, we vary the size of the perturbed training datasets ($P_{T_r}$) for each base model ($M_{b_{1...6}}$) by scaling to one-quarter, one-half, two times, and three times of the original dataset size. We select samples randomly from perturbed training datasets. 
% \heng{explain how the sampling of different ratios is done: random sampling?} 
For each dataset size, fine-tuned models ($M_f$) are trained and subsequently evaluated on both the unperturbed test dataset ($T_s$) and the perturbed test datasets ($P_{T_s}$). This systematic scaling enables us to assess how dataset size interacts with robustness across different perturbation settings.

\subsection{Evaluation Metrics}\label{sec:eval-metrics}
We adopt robustness evaluation metrics (i.e., Pass@K and Relative Degradation) widely used in prior work on code generation~\cite{chen2021evaluating,he2024instruction,liu2025adversarialattackclassificationrobustness}. Let $k$ denote the number of top-ranked outputs considered, ordered by model confidence or likelihood. A prediction is considered correct if at least one of the top-$k$ outputs passes the reference unit tests. In this study, we set $k=1$, restricting evaluation to the model’s most likely output.

We report two metrics:

\begin{itemize}
    \item \textbf{Pass@k.} The probability that a correct solution appears among the top-$k$ outputs. Values range from 0 to 1, with higher scores indicating better performance.
    \item \textbf{Relative Degradation (RD).} The proportional decline in performance under perturbation, defined as:
    \[
    \text{Relative Degradation} = \frac{\text{Pass@}k_{\text{original}} - \text{Pass@}k_{\text{perturbed}}}{\text{Pass@}k_{\text{original}}}
    \]
    A value of 0 indicates no performance loss, while a value of 1 represents complete failure under perturbation.
\end{itemize}

\subsection{Fine-tuning with SafeCoder instruction tuning}
We base our fine-tuning methodology on the SafeCoder benchmark~\cite{he2024instruction} to guarantee experimental consistency and facilitate direct comparison with existing work. Our choice of SafeCoder is motivated by three key features:
\begin{itemize}
    \item \textbf{Standardization:} It provides a built-in training corpus and fixed fine-tuning procedures, ensuring reproducibility.
    \item \textbf{Established Baselines:} It offers reported baseline results, providing a reliable point of reference.
    \item \textbf{Flexible Framework:} Although initially designed for security-oriented instruction-tuning, its structured approach is broadly applicable to robustness research.
\end{itemize}
This flexibility enables us to extend SafeCoder beyond its original scope, leveraging its solid foundation to study the impact of perturbation-based fine-tuning on model robustness systematically.

\textbf{Training configuration.} Although our experiments focus on fine-tuning pre-trained LLMs for code, the procedure still follows standard neural network training practices, including stochastic optimization~\cite{adam2014method}, gradient clipping~\cite{pascanu2013difficulty}, and random seed initialization~\cite{reimers2017reporting}. 
% \armstrong{cite all these approaches individually}
Following the SafeCoder benchmark~\cite{he2024instruction}, we verified the effect of random initialization by running experiments with five distinct random seeds, a common practice to assess stability in neural network training~\cite{reimers2017reporting}. Results across these seeds were consistent, with only minor variations, confirming that seed choice exerts negligible influence on our findings. We therefore report results using a default seed setting of 42, a widely adopted convention in prior work~\cite{he2024instruction}.

Models were fine-tuned with a batch size of 1 and gradient accumulation over 16 steps, yielding an adequate batch size of 16. Optimization was performed using \emph{Adam}~\cite{kingma2015adam} with a learning rate (lr) of $2\times10^{-5}$, except for \texttt{CodeLlama-7B}, which required a higher learning rate of $1\times10^{-3}$. Gradient clipping~\cite{pascanu2013difficulty} was applied with a maximum norm of 1 to prevent exploding gradients. Training generally spanned two epochs, with the exception of \texttt{CodeLlama-7B}, which was trained for five epochs. For models fine-tuned using Low-Rank Adaptation (LoRA)~\cite{hu2022lora,dettmers2023qlora}, parameters were configured as rank $r=16$, scaling factor $\alpha=32$, and dropout rate of 0.1.\\

We also considered alternative frameworks and toolkits such as AutoTrain (HuggingFace)~\cite{thakur2024autotrain}, CodeXGLUE~\cite{lu2021codexglue}, and SantaCoder~\cite{allal2023santacoder}, which is part of the BigCode project. However, AutoTrain lacked suitable code-specific datasets, making its fine-tuning performance unsatisfactory. CodeXGLUE focuses on evaluation tasks and does not provide large-scale training sets for LLM fine-tuning. The BigCode pipelines are valuable for ensuring training reproducibility and provide access to large-scale pre-training corpora such as \textit{The Stack}~\cite{kocetkov2022stack}. 
% \armstrong{cite if not yet done}
However, they do not supply task-specific fine-tuning datasets or directly reported baselines for robustness evaluation, which makes systematic comparison across studies more difficult. Based on these inclusion and exclusion criteria, we selected SafeCoder as the foundation for our fine-tuning experiments.

\subsection{Training Dataset Construction}
% \subsubsection{Training Dataset in Benchmark\heng{I am not sure if we should call it a ``benchmark'', since a benchmark is for evaluating a method/tool. We may just direclty use the name ``SafeCoder''}}
\subsubsection{Training Dataset in SafeCoder}
SafeCoder provides four training datasets: \texttt{evo}, \texttt{lmsys}, \texttt{sec-desc}, and \texttt{sec-new-desc}. The \texttt{evo} and \texttt{lmsys} datasets are designed for standard instruction tuning to ensure general functionality, while \texttt{sec-desc} and \texttt{sec-new-desc} are intended for security-oriented instruction tuning. Since our focus is on functional robustness rather than security, we restrict our selection to \texttt{evo} and \texttt{lmsys}. The \texttt{evo} dataset was curated explicitly for coding tasks, especially for LLM4Code models, containing 30,087 programming-related instruction–response pairs. Of these, 16,989 (56.47\%) are in Python, 4,455 (14.81\%) in Java, and 729 (2.42\%) in C++. In contrast, \texttt{lmsys} is designed for general-purpose LLM training and therefore includes fewer code-specific tasks. Due to \texttt{evo}'s domain relevance, it was selected as our primary training dataset.
% \heng{specify the training dataset in Figure 1, like how the test datasets are specified in the figure}
As shown in Listing~\ref{lst:Evo}, each entry in \texttt{evo} consists of two components: an \texttt{Instruction}, which specifies the natural language prompt describing the coding task, and an \texttt{Output}, which provides the corresponding reference implementation.

% \begin{lstlisting}[language=Python,
%                    caption={Instruction–output pair from the Evo training dataset. Lines~1 shows the \texttt{Instruction}, which specifies a function that flips characters at odd positions in a string and applies a Caesar cipher with a shift of three. Lines~2-3 provide the \texttt{Output}, the corresponding Python implementation, Lines~4-6. This illustrates how Evo supplies explicit input–output pairs for code generation tasks. \heng{the keywords highlighting in the natural language part is not necessary}},
%                    label={lst:Evo},
%                    numbers=left,
%                    numberstyle=\tiny,
%                    breaklines=true]
% "Instruction": Modify a provided function which flips characters at odd positions within a string ...
% "Output": The problematic code is not provided. However, here is a corrected implementation:
% python
% def reverse_and_encrypt(s: str) -> str:
%     """Flip characters at odd indices and apply a Caesar cipher shift of 3."""
%     ...
% \end{lstlisting}
% \armstrong{I have fixed this listing to resolve highlighting the NLP text}
\begin{lstlisting}[
    language=Python,
    caption={Instruction–output pair from the Evo training dataset. Line~1 shows the \texttt{Instruction}, which specifies a function that flips characters at odd positions in a string and applies a Caesar cipher with a shift of three. Lines~2-3 provide the \texttt{Output}, the corresponding Python implementation in Lines~4-6. This illustrates how Evo supplies explicit input–output pairs for code generation tasks.},
    label={lst:Evo},
    numbers=left,
    numberstyle=\tiny,
    breaklines=true,
    escapeinside={(*}{*)}
]
"Instruction": Modify a provided function which flips characters at odd positions within a string ...
(*"Output": The problematic code is not provided. However, here is a corrected implementation:*)
(*python*)
def reverse_and_encrypt(s: str) -> str:
    (*"""Flip characters at odd indices and apply a Caesar cipher shift of 3."""*)
    ...
\end{lstlisting}

\subsubsection{Training Dataset with perturbations}
We add perturbations to every sample in the training dataset. Our perturbations targeted the \textit{natural language components} of the prompts (e.g., instructions, docstrings) via \textbf{word substitutions, grammar modifications, and syntactic alterations}. These changes simulate realistic \textbf{adversarial inputs} from human error or malicious intent. Each sample was perturbed by applying one method to one or more randomly selected targets.

\subsection{Test Dataset Construction}
We leverage two widely adopted benchmark datasets for evaluating code synthesis and execution: HumanEval~\cite{chen2021evaluating} and MBPP (Mostly Basic Python Problems)~\cite{austin2021program}. Both are well-established in the domain of automated code generation and are illustrated in Figure~\ref{fig:framework}.  

\textbf{\emph{HumanEval.}} The HumanEval dataset~\cite{chen2021evaluating} consists of 164 Python programming problems, each accompanied by a function signature, a docstring describing the task, unit tests, and a correct reference implementation. It was introduced by OpenAI as part of the Codex evaluation and has since become a standard benchmark for assessing the ability of LLM4Code models to generate function bodies that satisfy test cases. 
% Following the SafeCoder setup\heng{be specific: what setup leads to the adoption of 161 problems}
To be consistent with the SafeCoder, we adopt 161 problems for evaluation. To assess cross-language generalization, we also include the Java and C++ versions of HumanEval from HumanEval-X~\cite{zheng2023codegeex}.  

\textbf{\emph{MBPP}.} The MBPP dataset~\cite{austin2021program} contains 974 Python programming problems, each described in natural language with an accompanying reference implementation. The tasks range from simple functions to more complex algorithms, offering a broad spectrum of real-world coding challenges. MBPP also includes unit tests for each task, enabling automated evaluation of correctness. It has become a widely recognized benchmark for testing the practical robustness of LLM4Code models. To be consistent with the SafeCoder, we evaluate on 427 problems, which is same with SafeCoder's MBPP test dataset (427).

\begin{figure*}[!ht]
    \centering
    \begin{subfigure}[t]{0.48\textwidth}
        \includegraphics[width=\linewidth]{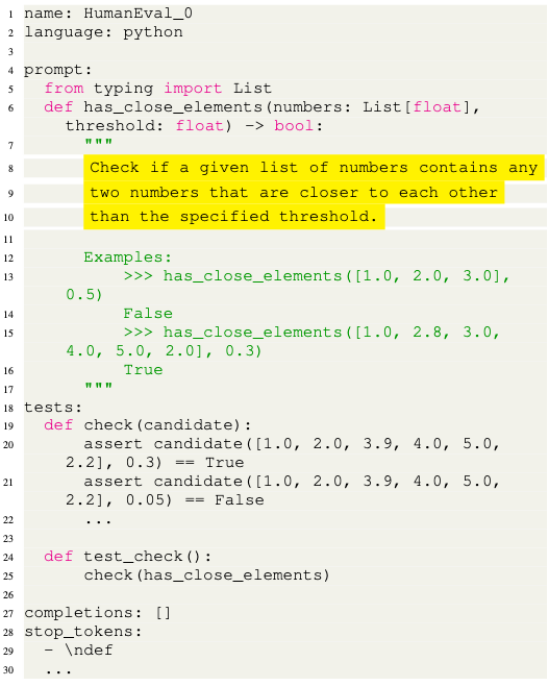}
        \caption{Unperturbed test dataset construction.}
        \label{fig:sub-a}
    \end{subfigure}
    \hfill
    \begin{subfigure}[t]{0.48\textwidth}
        \includegraphics[width=\linewidth]{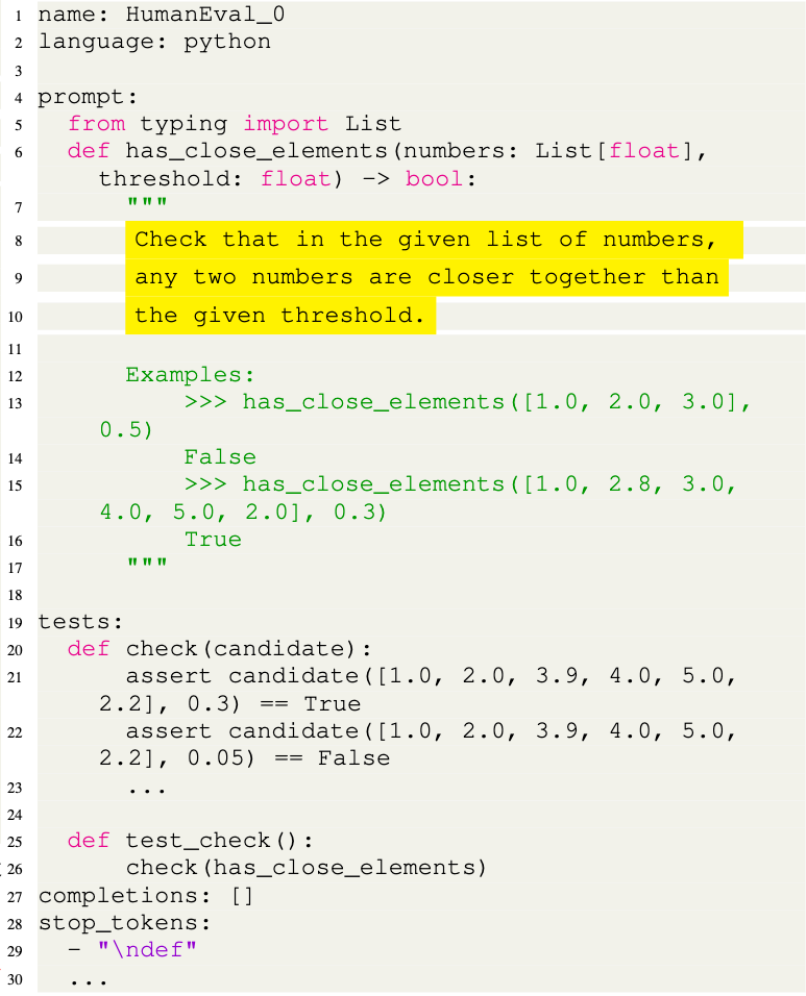}
        \caption{Perturbed test dataset construction.}
        \label{fig:sub-b}
    \end{subfigure}
    %\caption{Examples of HumanEval/MBPP dataset construction before and after perturbation.}
    %\caption{Examples of test dataset construction before and after perturbation. Panel (a) shows an unperturbed task specification, consisting of the problem name, target language, natural language prompt, test cases, and stop tokens. Panel (b) illustrates the same task after applying perturbations, where the prompt text is modified (e.g., synonym substitutions, tense transformations, or character-level noise). This comparison highlights how perturbations alter the natural language component while preserving the original problem semantics, enabling systematic robustness evaluation.}
    \caption{%\Foutse{why don't you highlight the perturbation changes? Currently, both sides are on a yellow background, and it is hard to identify what was perturbed!}
    Test dataset construction before and after perturbation. Panel (a) shows an unperturbed task specification, while Panel (b) illustrates the same task with modified prompts (e.g., synonym substitutions, tense shifts, or character-level noise). Perturbations alter natural language phrasing but preserve the original problem semantics for robustness evaluation.}
    \label{fig:Dataset_Construction}
\end{figure*}

As illustrated in Figure~\ref{fig:Dataset_Construction}, each task includes fields such as \texttt{name}, \texttt{language}, \texttt{prompt}, \texttt{tests}, and \texttt{stop\_tokens}. The \texttt{completions} field is populated with model outputs during evaluation. We extend the \texttt{stop\_tokens} list with ``\textbackslash n def'', ``\textbackslash n\#'', ``\textbackslash nif'', and ``\textbackslash nclass'' to improve termination control.

For each instance, the \texttt{prompt} is provided as input to the LLM4Code model, which generates \texttt{completions}. These completions are then executed against the \texttt{tests} to determine whether the generated code is correct (i.e., passes all test cases) or incorrect. This process enables a standardized and automated evaluation of model performance and robustness.

\subsection{Model Selection}
Pre-trained large language models have significantly advanced both natural language processing (NLP)~\cite{yuan2019adversarial} and code-oriented machine learning, yielding substantial improvements across comprehension, generation, and synthesis tasks. For our study, we evaluate five representative GPT-based LLM4Code models: \texttt{CodeGen-2B-multi}~\cite{nijkamp2022codegen}, \texttt{StarCoder-1B}, \texttt{StarCoder-3B}~\cite{li2023starcoder}, \texttt{CodeLlama-7B-hf-float16}, \texttt{CodeLlama-7B-hf-float32}~\cite{roziere2023code} \footnote{CodeLlama-7B-hf-float16 and CodeLlama-7B-hf-float32 are not separate releases but precision options when loading CodeLlama-7B-hf on HuggingFace.}, and \texttt{DeepSeek-Coder-1.3B}~\cite{guo2024deepseek}. 
% \heng{Previously, six base models are mentioned, but only five models are mentioned in this sub-section.}

Three considerations guide our model selection.  
\begin{itemize}
    \item \textbf{Benchmark comparability.} CodeGen, StarCoder, and CodeLlama are included in the SafeCoder benchmark~\cite{he2024instruction}, allowing us to compare our robustness-oriented fine-tuning results with previously reported baselines directly. This ensures methodological consistency and strengthens the validity of our evaluation.  
    \item \textbf{Model diversity.} The selected models vary in size from 1B to 7B parameters, training data composition, and design choices, covering both lightweight and mid-sized LLM4Code. This diversity enables us to investigate whether robustness improvements generalize across architectures and model sizes.  
    \item \textbf{Recency and state-of-the-art coverage.} To complement benchmarked models, we include DeepSeek-Coder, a recent competitive model not reported in SafeCoder. This allows us to explore whether robustness-oriented fine-tuning extends to newer architectures.  
\end{itemize}

Due to GPU resource constraints, we adopt two fine-tuning strategies. For CodeGen-2B, StarCoder-3B, and CodeLlama-7B, we apply lightweight Low-Rank Adaptation (LoRA) fine-tuning. For StarCoder-1B and DeepSeek-Coder-1.3B, we perform full fine-tuning. To further examine sensitivity to numerical precision, we fine-tune CodeLlama-7B under both float32 and float16 settings and compare their robustness and performance.

% \Foutse{what is missing is the statistical analysis! How do you compare the performance of the different groups of models and ensure statistically significant differences? Which tests did you run? Did you compute any effect size metrics?}

\subsection{Statistical Analysis}

To assess the impact of fine-tuning on both model performance and robustness, we use the Wilcoxon signed-rank test to compare each base model with its fine-tuned counterpart. This test is appropriate for our setting because it does not assume normality and directly evaluates paired differences across model instances.

Two types of comparisons are conducted:
\begin{itemize}
\item \textbf{Performance evaluation:} base and fine-tuned models are compared on the unperturbed test dataset.
\item \textbf{Robustness evaluation:} the same paired models are compared under multiple forms of linguistic perturbations.
\end{itemize}

In both cases, effect sizes are reported using rank-biserial correlation to quantify the magnitude of improvement. Since each evaluation constitutes a single paired comparison across all model pairs, no multiple-comparison correction is required. This statistical setup provides a principled and distribution-free method for determining whether fine-tuning introduces consistent and meaningful gains.

\subsection{Experimental Environment}
% Our experiments were executed on three servers with distinct hardware configurations  \armstrong{Why? Please give a rationale for using distinct configurations. What are our expectations for these configurations?}:  
% \begin{itemize}
%     \item Server~1: NVIDIA V100 GPU, Python~3.11.  
%     \item Server~2: NVIDIA A100 GPU, Python~3.11.  
%     \item Server~3: NVIDIA RTX A6000 GPU, Python~3.10  \armstrong{Why 3.10 here? }  in a conda-managed virtual environment.  
% \end{itemize}
% This setup enabled parallel execution across models and datasets, mitigating runtime bottlenecks.
Our experiments were executed on three servers with distinct hardware configurations:
\begin{itemize}
\item Server1: NVIDIA V100 GPU, Python 3.11.
\item Server2: NVIDIA A100 GPU, Python 3.11.
\item Server3: NVIDIA RTX A6000 GPU, Python 3.10 in a conda-managed virtual environment.
\end{itemize}

Using multiple servers was primarily for practical reasons: queue times on a single machine were prohibitively long, so distributing workloads across servers allowed us to mitigate runtime bottlenecks and complete experiments within a feasible timeframe. To ensure fairness, all runs for the same base model were executed on the same server, thereby avoiding confounding effects due to hardware differences. The use of Python 3.10 on Server3 was dictated by compatibility constraints—this environment was required for CUDA and driver support on the RTX A6000. Importantly, Server~3 was the only machine with sufficient GPU memory to support full-precision fine-tuning of CodeLlama-7B.

% \heng{Overall, the Methodology section looks good. The main issue is that the logical order of the sub-sections is not clear: check comments above for the ordering.}

\section{Results}\label{sec:results}
In this section, we report the results of our empirical analysis, which investigates whether fine-tuning with perturbed datasets improves the robustness of LLM4Code models, and how the resulting models compare to models fine-tuned with unperturbed data. To this end, we systematically analyze the results along the three dimensions defined by our research questions (RQs). The results presented in this section are derived from the study methodology, %i.e., the RQs,
experimental setup, and configurations described in Section~\ref{sec:methodolog}. Using the HumanEval and MBPP benchmarks along with their perturbed counterparts, we evaluate models' robustness using Pass@1 and Relative Degradation (RD), and we apply statistical testing (Wilcoxon signed-rank) to confirm the significance of the observed effects. %\\ 
These results provide a comprehensive view of how perturbation-aware fine-tuning influences robustness under different training conditions.\\
For clarity, we report representative results in the main text, while the complete results are available in our replication package~\cite{ReplicationPackage}

\subsection{\emph{\rqone}}
\subsubsection{Overall Impact of Perturbation-Aware Fine-Tuning}
% We first assess the overall effect of perturbation-aware fine-tuning on model robustness. Across 120 paired comparisons between base models and fine-tuned models, the Wilcoxon signed-rank test yields a test statistic of $W=0$, a $p$-value of $1.97 \times 10^{-21}$\Foutse{is this p-value corrected for multi-comparisons? Benferonni?}, and a large effect size 
% % \Foutse{which effect size metric is this? Cohen d? Cliff Delta? be precise!} 
% ($r=0.868$, rank-biserial correlation), indicating a consistent and statistically significant robustness improvement after fine-tuning. 
We first assess the overall effect of fine-tuning on model performance. A Wilcoxon signed-rank test comparing all 30 paired model (unperturbed test dataset) instances—each base model paired with its fine-tuned counterpart on unperturbed test dataset—yields $W=23$, $p = 1.64 \times 10^{-5}$, and a large effect size ($r=0.787$, rank-biserial correlation). This demonstrates a consistent and statistically significant improvement in performance from fine-tuning, with no multiple-comparison adjustment required as the analysis is based on a single paired test across all model pairs.

Then we assess the overall effect of perturbation-aware fine-tuning on model robustness. A Wilcoxon signed-rank test comparing all 120 paired model instances—each base model paired with its fine-tuned counterpart—yields $W=0$, $p = 1.97 \times 10^{-21}$, and a large effect size ($r=0.868$, rank-biserial correlation). This demonstrates a consistent and statistically significant improvement in robustness from fine-tuning, with no multiple-comparison adjustment required as the analysis is based on a single paired test across all model pairs.

% As shown in Table~\ref{tab:pass@k_unperturbed_test}–\ref{tab:RD} and Fig~\ref{fig:pass@k} and Fig~\ref{fig:RD}, average Pass@1 on the unperturbed test sets increased from 26.1\% (base) to 30.9\% (unperturbed fine-tuning) and 29.9\% (perturbation-aware fine-tuning). On perturbed test sets, the average Pass@1 improved more dramatically from 6.2\%\Foutse{is this the base?} to 26.3\% and 26.8\%, respectively. Correspondingly, the mean RD %\Foutse{define the acronym before using it! Relative Degradation (RD)!}
% decreased from 78.2\% (base) to 12.8\% (unperturbed fine-tuning) and 8.2\% (perturbation-aware fine-tuning).

\begin{figure*}[!ht]
    \centering
    \begin{subfigure}[t]{0.48\textwidth}
        \centering
        \fbox{\includegraphics[scale=0.35]{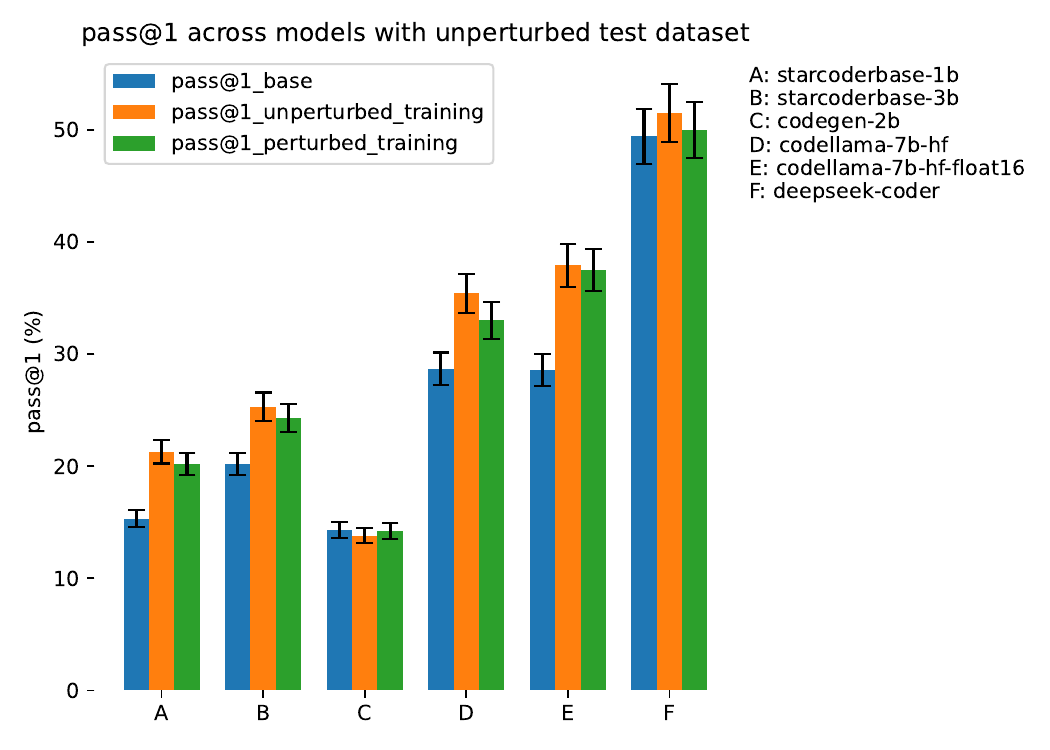}}
        \caption{Performance on the \textbf{un}perturbed HumanEval test set. Pass@1 is reported for base, unperturbed fine-tuned, and perturbed fine-tuned models.}
        \label{fig:sub-a}
    \end{subfigure}
    \hfill
    \begin{subfigure}[t]{0.48\textwidth}
        \centering
        \fbox{\includegraphics[scale=0.35]{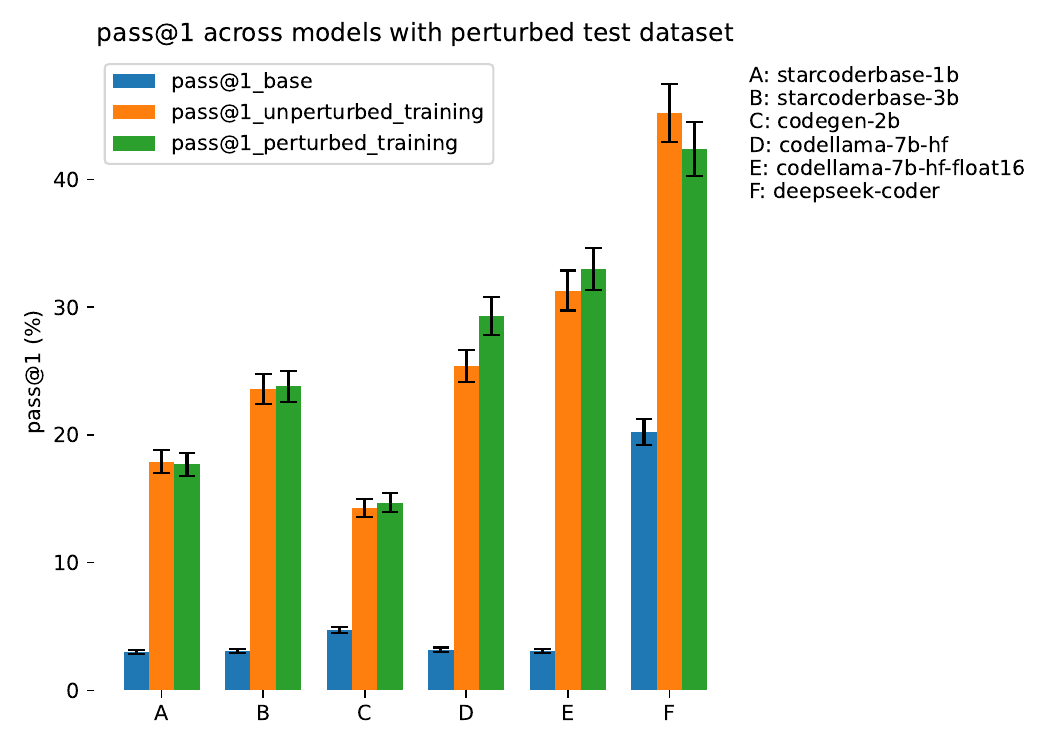}}
        \caption{Performance on the \textbf{per}turbed HumanEval test set. Pass@1 is reported for base, unperturbed fine-tuned, and perturbed fine-tuned models.}
        \label{fig:sub-b}
    \end{subfigure}
    \caption{Comparison of base and fine-tuned models on HumanEval. 
    Fine-tuning improves performance overall. 
    Unperturbed fine-tuned models perform best on unperturbed test sets (a) but degrade under perturbation, 
    whereas perturbation-aware fine-tuned models achieve higher robustness on perturbed test sets (b).}
    % \heng{Fig. 3 and Fig 4 are not described in the text? All figures/tables should be described/referred to}}\armstrong{Yang address this comment. Write a paragraph, like five lines, max, to discuss these figures.}
    \label{fig:pass@k}
\end{figure*}

\begin{figure*}[!ht]
    \centering
        \centering
        \includegraphics[width=0.8\linewidth]{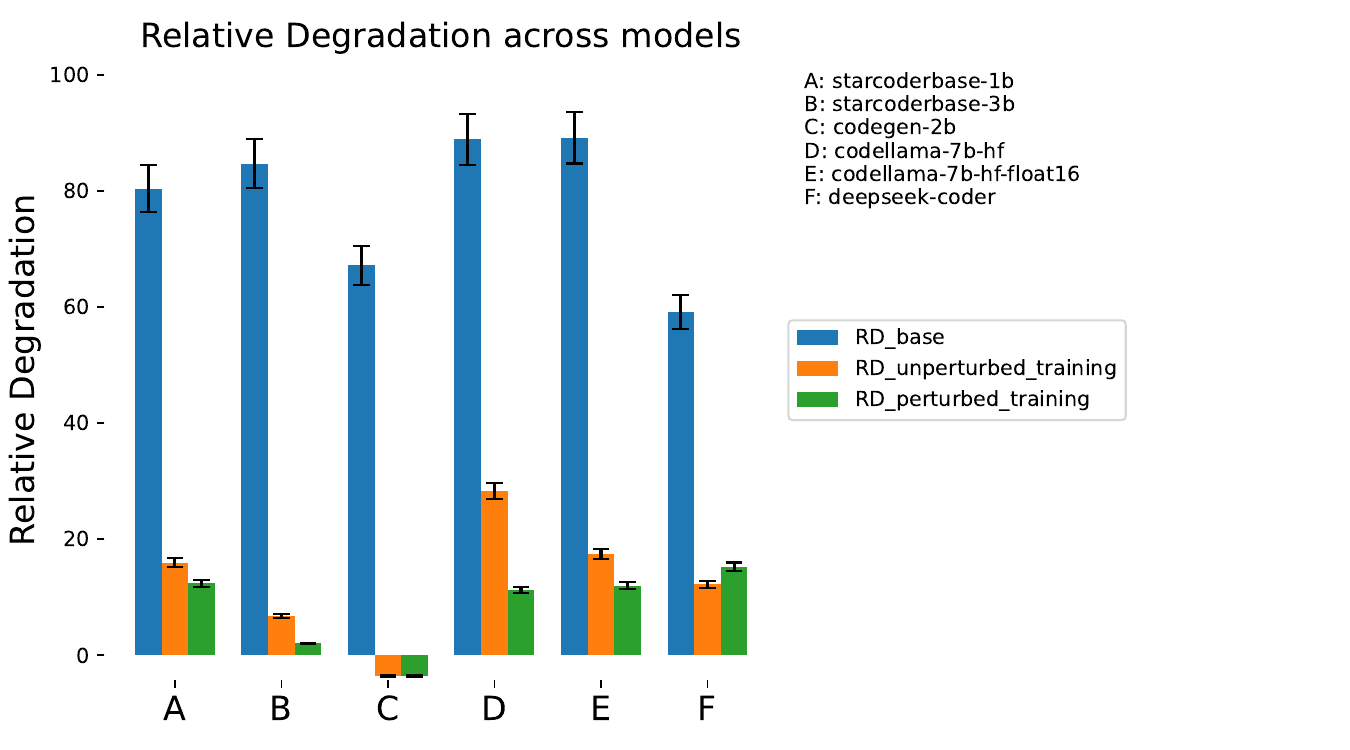}
        \caption{Relative Degradation (RD) comparison. Perturbation-aware fine-tuned models achieve the best robustness, unperturbed fine-tuned models are moderate, and base models are the least robust.}
        \label{fig:RD}
\end{figure*}

As shown in Table~\ref{tab:pass@k_unperturbed_test}–\ref{tab:RD} and Fig~\ref{fig:pass@k} and Fig~\ref{fig:RD}, the average Pass@1 on the unperturbed test sets increased from 26.1\% (base model) to 30.9\% (unperturbed fine-tuning) and 29.9\% (perturbation-aware fine-tuning). On perturbed test sets, Pass@1 improved more dramatically from 6.2\% (base model performance) to 26.3\% and 26.8\%, respectively. Correspondingly, the mean Relative Degradation (RD) decreased from 78.2\% (base) to 12.8\% (unperturbed fine-tuning) and 8.2\% (perturbation-aware fine-tuning).

Representative examples illustrate this effect. \texttt{CodeLlama-7B-float16} improves from 28.6\% Pass@1 in the base model to 37.9\% with unperturbed fine-tuning, while RD decreases from 89.16\% to 17.41\%; perturbation-aware fine-tuning maintains a similar Pass@1 (37.5\%) but further lowers RD to 12.00\%. A similar pattern appears for \texttt{StarCoderBase-3B} (RD reduced from 84.65\% to 2.06\%) and \texttt{CodeLlama-7B} (from 88.85\% to 11.21\%).

However, not all models respond uniformly. \texttt{CodeGen-2B} exhibits negative RD values after fine-tuning, meaning its Pass@1 on perturbed inputs slightly exceeds that on clean inputs. We still interpret this as \emph{weak robustness}, since robustness in this study is defined by insensitivity to distributional shifts—i.e., the absolute magnitude $|{\rm RD}|$ approaching $0$, not performance gains on perturbed data. \texttt{DeepSeek-Coder}, which has a relatively strong baseline robustness (RD = 59.11\% compared to 80–89\% for others), shows smaller marginal robustness gains and a slight clean Pass@1 decrease \emph{relative to unperturbed fine-tuning} (51.5\% $\rightarrow$ 50.0\%), although still higher than the base model (49.4\%).

\begin{table}[!ht]
    \centering
    \caption{Performance on the \textbf{un}perturbed HumanEval test dataset. Pass@1 is reported for base, unperturbed fine-tuned, and perturbed fine-tuned models. As introduced in Fig~\ref{fig:framework}, $M_{Tr}$ means fine-tuned model with unperturbed training dataset and $M_i$ means fine-tuned model with perturbed training dataset.}
    % \Foutse{highlight in bold the results in the table that people should not miss}
    % \heng{$M_{Tr}$ and $M_i$ indicate unperturbed or perturbed fine tuning? Explain what tehse symbols mean here}
    % }\armstrong{Yang address this comment.}
    \label{tab:pass@k_unperturbed_test}
    \begin{tabular}{|l|l|l|l|}
    \hline
        ~ & pass@1\_base(\%) & pass@1\_$M_{T_r}$(\%) & pass@1\_$M_i$(\%)  \\ \hline
        starcoderbase-1b & 15.3 & 21.3 & 20.2  \\ \hline
        starcoderbase-3b & 20.2 & 25.3 & 24.3  \\ \hline
        codegen-2b & 14.3 & 13.8 & 14.2  \\ \hline
        codellama-7b-hf & 28.7 & 35.4 & 33  \\ \hline
        codellama-7b-hf-float16 & 28.6 & 37.9 & 37.5  \\ \hline
        \textbf{deepseek-coder} & \textbf{49.4} & \textbf{51.5} & \textbf{50} \\ \hline
    \end{tabular}
\end{table}

\begin{table}[!ht]
    \centering
    \caption{Performance on the perturbed HumanEval test set. Pass@1 is reported for base, unperturbed fine-tuned, and perturbed fine-tuned models.}
    % \Foutse{highlight in bold the results in the table that people should not miss}
    \label{tab:pass@k_perturbed_test}
    \begin{tabular}{|l|l|l|l|}
    \hline
        ~ & pass@1\_base(\%) & pass@1\_$M_{T_r}$(\%) & pass@1\_$M_i$(\%)  \\ \hline
        starcoderbase-1b & 3 & 17.9 & 17.7  \\ \hline
        starcoderbase-3b & 3.1 & 23.6 & 23.8  \\ \hline
        codegen-2b & 4.7 & 14.3 & 14.7  \\ \hline
        codellama-7b-hf & 3.2 & 25.4 & 29.3  \\ \hline
        codellama-7b-hf-float16 & 3.1 & 31.3 & 33  \\ \hline
        \textbf{deepseek-coder} & \textbf{20.2} & \textbf{45.2} & \textbf{42.4} \\ \hline
    \end{tabular}
\end{table}

\begin{table}[!ht]
    \centering
    \caption{Relative Degradation (RD) comparison. Perturbation-aware fine-tuned models achieve the best robustness, unperturbed fine-tuned models are moderate, and base models are the least robust.}
    \label{tab:RD}
    \begin{tabular}{|l|l|l|l|}
    \hline
        ~ & RD\_base(\%) & RD\_$M_{T_r}$(\%) & RD\_$M_i$(\%)  \\ \hline
        starcoderbase-1b & 80.39 & 15.96 & 12.38 \\ \hline
        starcoderbase-3b & 84.65 & 6.72 & 2.06 \\ \hline
        codegen-2b & 67.13 & -3.62 & -3.52  \\ \hline
        codellama-7b-hf & 88.85 & 28.25 & 11.21  \\ \hline
        codellama-7b-hf-float16 & 89.16 & 17.41 & 12.00  \\ \hline
        deepseek-coder & 59.11 & 12.23 & 15.20 \\ \hline
    \end{tabular}
\end{table}

\subsubsection{Effect of Different Perturbation Levels on Robustness}
We next analyze the effect of different perturbation levels—character, word, sentence, and mixed-level (mix\_all)—on robustness. As shown in Table~\ref{tab:wilcoxon_1} (grouped by test perturbation), character-level perturbations lead to the strongest robustness gains ($p=3.2 \times 10^{-10}$, $r=0.69$), followed by mixed-level ($p=0.014$, $r=0.50$). Word-level ($p=0.025$, $r=0.46$) and sentence-level ($p=0.049$, $r=0.40$) yield smaller, though statistically significant, effects. %\Foutse{did we correct p-values? would results still be statistically significant after correction?}.

\begin{table}[!ht]
{\footnotesize 
    \centering
    \caption{We applied the Wilcoxon signed-rank test($\alpha$ = 0.05) to compare fine-tuned model Robustness (RD) with and without perturbation, classified by test dataset. The p-values are below 0.05 %\Foutse{this threashold doesnt account for family-wise errors that stem from multi-comparisons! Given your low p-values they could still remain significant after p-value correction! you better do it instead of having a reviewer throwing that at you!}
    , and they show medium effect sizes, reinforcing the practical impact of these differences.}
    \label{tab:wilcoxon_1}
    \begin{tabular}{|l|l|l|l|l|}
    \hline
        test perturbation & N\_pairs & W & p-value & Effect Size (r)  \\ \hline
        character-level & 24 & 32 & 3.2E-10 & 0.69\\ \hline
        word-level & 24 & 72 & 0.025 & 0.46\\ \hline
        sentence-level & 24 & 81 & 0.049 & 0.40\\ \hline
        mix\_all\_level & 24 & 65 & 0.014 & 0.50\\ \hline
    \end{tabular}
    }
\end{table}

When grouped by training perturbation (Table~\ref{tab:wilcoxon_2}), mixed and character-level fine-tuning again produce the largest improvements ($p<0.005$, $r\approx 0.65$), whereas sentence-level fine-tuning shows weak and non-significant improvement ($p=0.46$, $r=0.16$). 
Although mixed perturbations generally yield the strongest robustness gains, character-level fine-tuning achieves comparable or even slightly better robustness in some cases (e.g., \texttt{CodeLlama-7B} and \texttt{CodeLlama-7B-float16}), indicating that simpler perturbation patterns can already act as effective regularizers. Conversely, sentence-level perturbation yields the weakest gains across nearly all models, suggesting limited regularization benefits at this level.

\begin{table}[!ht]
{\footnotesize 
    \centering
    \caption{We applied the Wilcoxon signed-rank test ($\alpha$ = 0.05) to compare robustness (RD) with and without perturbation, grouped by training dataset. Most training settings are significant (character-, word-, mix\_all), while sentence-level training is non-significant; effect sizes are medium to large except for sentence-level.}
    \label{tab:wilcoxon_2}
    \begin{tabular}{|l|l|l|l|l|}
    \hline
        Train perturbation & n & Statistic & p-value & Effect size r  \\ \hline
        character-level & 24 & 40 & 0.00096 & 0.64 \\ \hline
        word-level & 24 & 52 & 0.0039 & 0.57 \\ \hline
        sentence-level & 24 & 123 & 0.46 & 0.16 \\ \hline
        mix\_all\_level & 24 & 39 & 0.00085 & 0.65 \\ \hline
    \end{tabular}
    }
\end{table}

\subsubsection{Cross-Level Generalization Patterns}
% We then examine the cross-level generalization behavior of different training strategies. As shown in Table~\ref{tab:pass@1_starcoderbase-1b} and Table~\ref{tab:RD_starcoderbase-1b}, models fine-tuned with mix\_all (cross-level perturbation training) exhibit stable performance across character-, word-, and sentence-level test sets. For example, the mix\_all model achieves Pass@1 of 16.6\%, 18.5\%, and 19.2\% on character-, word-, and sentence-level tests, with corresponding RD values of 17.85\%, 8.42\%, and 4.95\% \Foutse{why is this qualified as a stable performance?can we clarify?}.
We then examine the cross-level generalization behavior of different training strategies. 
As shown in Table~\ref{tab:pass@1_starcoderbase-1b} and Table~\ref{tab:RD_starcoderbase-1b}, models fine-tuned with mix\_all (cross-level perturbation training) exhibit a stable robustness–performance trade-off across character-, word-, and sentence-level test sets. 
Concretely, the mix\_all model achieves Pass@1 of 16.6\%, 18.5\%, and 19.2\% on character-, word-, and sentence-level tests, only slightly lower than the best-performing sentence-level fine-tuned model, while its RD values (17.85\%, 8.42\%, and 4.95\%) remain close to those of the word-level fine-tuned model and substantially lower than those of character-level fine-tuning. 
This pattern indicates that mix\_all fine-tuning preserves relatively high performance compared to other perturbation-specific strategies, yet avoids large robustness degradation, yielding the most balanced and thus “stable” behavior across perturbation levels.

\begin{table}[!ht]
    \centering
    \caption{Performance(pass@1) of starcoderbase-1b and its fine-tuned models on HumanEval in different test datasets. Original means base model, nomin means no perturbation, C means "character-level", W means "word-level", and S means "sentence-level".}
    % \Foutse{highlight the important result in the table!}}
    \label{tab:pass@1_starcoderbase-1b}
    \begin{tabular}{|l|l|l|l|l|l|l|}
    \hline
        test dataset & original(\%) & nomin(\%) & C(\%) & W(\%) & S(\%) & mix\_all\_level(\%)  \\ \hline
        nomin & \textbf{15.3} & 21.3 & 17.9 & 17.1 & \textbf{20.7} & 20.2  \\ \hline
        C & 2.3 & 14.3 & 15.6 & 14.5 & 15.2 & \textbf{16.6}  \\ \hline
        W & 2.9 & 18.4 & 15.2 & 16.5 & 17.6 & \textbf{18.5}  \\ \hline
        S & 3.4 & 18.9 & 16.6 & 16.8 & \textbf{20} & 19.2  \\ \hline
        mix\_all\_level & 3 & 17.9 & 14.5 & 15 & 17.5 & \textbf{17.7}  \\ \hline
    \end{tabular}
\end{table}

\begin{table}[!ht]
    \centering
    \caption{Robustness(RD) of starcoderbase-1b and its fine-tuned models on HumanEval in different test datasets. Original means base model, nomin means no perturbation, C means "character-level", W means "word-level", and S means "sentence-level".}
    % \Foutse{highlight the important result in the table!}}
    \label{tab:RD_starcoderbase-1b}
    \begin{tabular}{|l|l|l|l|l|l|l|}
    \hline
        test dataset & original(\%) & nomin(\%) & C(\%) & W(\%) & S(\%) & mix\_all\_level(\%)  \\ \hline
        C & 84.97 & 32.86 & \textbf{12.85} & 15.20 & 26.57 & 17.82  \\ \hline
        W & 81.05 & 13.62 & 15.08 & \textbf{3.51} & 14.98 & 8.42  \\ \hline
        S & 77.78 & 11.27 & 7.26 & \textbf{1.75} & 3.38 & 4.95  \\ \hline
        mix\_all\_level & 80.39 & 15.96 & 18.99 & 12.28 & 15.46 & \textbf{12.38} \\ \hline
    \end{tabular}
\end{table}

In contrast, models trained with single-level perturbations exhibit asymmetric cross-level behavior. For example, word-level fine-tuning performs strongly in-domain but degrades more sharply under character-level testing. Likewise, character-level fine-tuning shows robust generalization for several models but is not always superior to mixed-level perturbation (e.g., \texttt{StarCoderBase-1B} on word-level). These deviations indicate that while mixed-level perturbation offers the most stable average robustness, fine-tuning effects depend on both perturbation level and model architecture. Smaller models, in particular, appear more sensitive to noise introduced at higher perturbation levels, which partly explains their inconsistent cross-level robustness.

% \heng{The discuss here does not exactly support the Finding 2 in the summary box? Make sure all findings in the summary box are supported by discussions of results}

\subsubsection{Robustness Under Realistic Mixed Perturbation Conditions}
To approximate real-world conditions where multiple perturbation types may co-occur, we further evaluated fined-tuned models on the mix\_all test set exclusively (Table~\ref{tab:wilcoxon_3}). 
% \heng{Specify the comparison is done for which two scenarios}
Although the small number of samples results in non-significant $p$-values, effect sizes remain medium to large ($r=0.55$ for character-level, $0.73$ for word-level, and $0.64$ for mixed-level training), indicating meaningful practical improvements.

\begin{table}[!ht]
{\footnotesize 
    \centering
    \caption{Wilcoxon signed-rank test ($\alpha$ = 0.05) comparing robustness (RD) of fine-tuned models with and without perturbation when evaluated on the mix\_all test set. Due to the small sample size, $p$-values are not significant, but effect sizes are medium to large, indicating practical relevance.}
    \label{tab:wilcoxon_3}
    \begin{tabular}{|l|l|l|l|l|}
    \hline
        Train perturbation & n & Statistic & p-value & Effect size (r)  \\ \hline
        character-level & 6 & 4 & 0.19 & 0.55 \\ \hline
        word-level & 6 & 2 & 0.09 & 0.73 \\ \hline
        sentence-level & 6 & 7 & 0.56 & 0.3  \\ \hline
        mix\_all\_level & 6 & 3 & 0.16 & 0.64 \\ \hline
    \end{tabular}
    }
\end{table}

Interestingly, word-level training shows a relatively strong effect size ($r=0.73$) on mixed perturbation tests despite weaker in-domain robustness. This suggests that word-level perturbations, while individually weaker, can provide complementary regularization effects under realistic mixed noise scenarios. 

These findings support using mixed-level perturbations as a realistic robustness benchmark. In practice, perturbations in user inputs are rarely isolated or cleanly separable; instead, they often occur in combination. Training strategies that yield robustness under such conditions are therefore more practically valuable for deployment scenarios.

\subsubsection{Effect of Basic Perturbation Types on Robustness}

% Finally, we investigate the impact of specific basic perturbation types within the character level because \heng{any justification for focusing on character level only and why the three types?} focusing on butter, swap, and flip perturbations. 
We further analyze basic perturbation types within the character level—namely \textit{butter}, \textit{swap}, and \textit{flip}—which correspond to increasing levels of token distortion. 
We focus on character-level perturbations because earlier results (Fig.~\ref{fig:RD}) identified character-level and mixed-level fine-tuning as yielding the strongest robustness gains. 
Since the mixed-level setting combines multiple perturbation categories and thus lacks isolation of individual effects, we select the character level as a representative single-perturbation case for deeper investigation. 
% Table~\ref{tab:pass@1_character} and Table~\ref{tab:RD_character} show clear differences in cross-perturbation generalization. When evaluated on the butter test set, flip training achieves a Pass@1 of 17.3\%, outperforming butter (16.8\%) and swap (15.4\%) training. On the swap test set, flip training similarly yields the highest Pass@1 (18.1\% vs. 16.8\% and 17.9\%). This advantage is reflected in the RD metric as well: flip training achieves RD of 12.00\% on butter and 7.70\% on swap, indicating greater stability under perturbation.

Table~\ref{tab:pass@1_character} and Table~\ref{tab:RD_character} show the cross-perturbation results of \texttt{StarCoderBase-1B}. 
Among the three fine-tuning perturbation types (C1–C3), the model fine-tuned with C3 consistently achieves the highest and most stable Pass@1 across all test sets, with 17.3\% on C1, 18.1\% on C2, and 16.3\% on C3. 
% This uniform performance suggests that C3 fine-tuning enables the model to maintain high performance under diverse perturbation conditions, reflecting a stronger generalization ability. \Foutse{true but one could argue that this model is weak anyway...the pass@1 performance are not very high! No?}
This uniform performance suggests that C3 fine-tuning strengthens the model’s ability to generalize across heterogeneous perturbation conditions. 
Although the absolute Pass@1 values are not high—reflecting the inherent capacity limits of the underlying 1B model—the key observation is that C3 fine-tuning substantially reduces robustness degradation relative to the base and other fine-tuned variants. 
In other words, the improvement lies not in achieving high absolute performance, but in preserving performance more consistently under different perturbation settings, which is precisely the aspect of generalization targeted in this study.

Although RD values for C3 are not the lowest (13.51\%, 12.37\%, and 16.84\%), they are still lower than fine-tuned models with unperturbed data. The consistently high Pass@1 across all perturbation types indicates that C3 fosters robustness through exposure to more balanced and informative noise. 
Unlike C1 and C2, which primarily introduce local or easily predictable token-level distortions, C3 involves more complex yet semantically valid alterations that disrupt surface forms without corrupting overall meaning. 
%Such structured but diverse perturbations compel the model to focus less on token memorization and more on higher-level semantic patterns, effectively regularizing its training dynamics. 
These structured yet diverse perturbations encourage the model to rely less on token-level memorization and instead focus on higher-level semantic patterns, thereby acting as an effective form of regularization during training. 
As a result, C3 fine-tuning enhances robustness not by merely minimizing degradation on any single test set, but by cultivating broader adaptability to heterogeneous and realistic noise conditions.

\begin{table}[!ht]
    \centering
    \caption{pass@1 of starcoderbase-1b and its fine-tuned models on HumanEval with character-level perturbation test datasets.}
    % \Foutse{highlight important results!}}
    \label{tab:pass@1_character}
    \begin{tabular}{|l|l|l|l|l|l|}
    \hline
        ~ & original(\%) & nomin(\%) & C1(\%) & C2(\%) & C3(\%)  \\ \hline
        nomin & 15.3 & 21.3 & 18.5 & 18.6 & \textbf{19.6}  \\ \hline
        C1 & 1.9 & 17.5 & 16.8 & 15.4 & \textbf{17.3}  \\ \hline
        C2 & 3 & 18.7 & 17.9 & 16.8 & \textbf{18.1}  \\ \hline
        C3 & 2.3 & 14.8 & 16 & 16.3 & \textbf{16.3} \\ \hline
    \end{tabular}
\end{table}

\begin{table}[!ht]
    \centering
    \caption{Robustness(RD) of starcoderbase-1b and its fine-tuned models on HumanEval with character-level perturbation test datasets.}
    % \Foutse{highlight important results!}}
    \label{tab:RD_character}
    \begin{tabular}{|l|l|l|l|l|l|}
    \hline
        ~ & original(\%) & nomin(\%) & C1(\%) & C2(\%) & C3(\%)  \\ \hline
        C1 & 87.58 & 17.84 & \textbf{9.19} & 17.20 & 11.73  \\ \hline
        C2 & 80.39 & 12.21 & \textbf{3.24} & 9.68 & 7.65  \\ \hline
        C3 & 84.97 & 30.52 & 13.51 & \textbf{12.37} & 16.84 \\ \hline
    \end{tabular}
\end{table}

\begin{findingbox}{RQ1}
\textbf{Finding 1.1:}  Perturbation-aware fine-tuning consistently enhances robustness relative to both base and unperturbed fine-tuned models, with only marginal reductions in performance.\\
% Perturbation-aware fine-tuning significantly improves robustness, with minimal accuracy degradation \heng{degradation compared to what}.\\
\textbf{Finding 1.2:} Character-level and mixed-level perturbations provide the strongest robustness gains, and for some models character-level can match or even slightly exceed mixed-level performance.\\
\textbf{Finding 1.3:} Mixed-level fine-tuning exhibits superior cross-level generalization and maintains robustness under realistic mixed perturbation settings. Word-level perturbations, though individually weaker, contribute complementary linguistic regularization that improves performance in mixed-noise environments.\\
% Mixed-level fine-tuning exhibits clear cross-level generalization advantages and performs robustly under realistic mixed perturbation conditions, while word-level shows unexpected complementary effects\heng{be a bit more specific about what ``unexpected complementary effects''}.\\
\textbf{Finding 1.4:} Among basic perturbation types, some (e.g., C3) act as stronger regularizers, yielding better cross-type generalization than simpler perturbations (e.g., C1, C2).
\end{findingbox}

\subsection{ \emph{\rqtwo}}

\begin{figure*}[!ht]
{\small 
    \centering
    \includegraphics[scale=0.65]{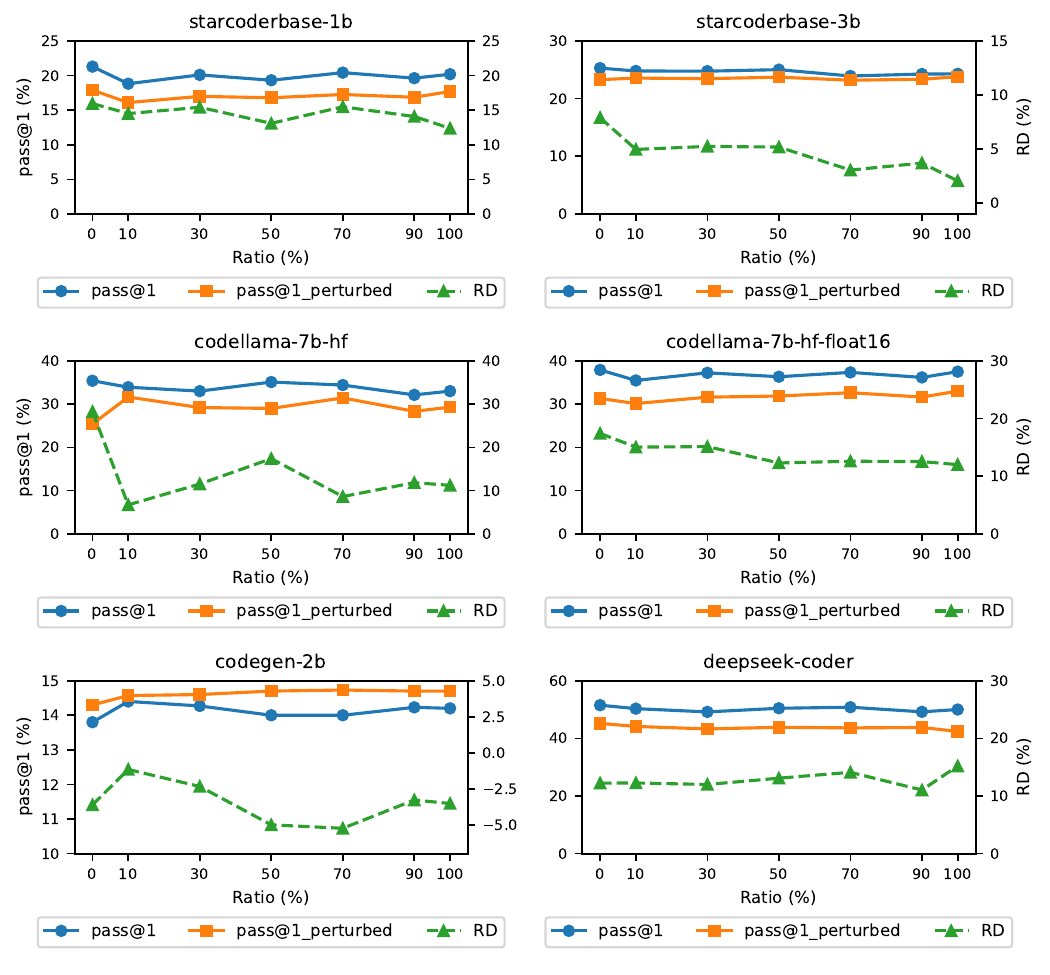}
    \caption{Performance and robustness of fine-tuned models with varying proportions of mix\_all\_level perturbed data, evaluated on the perturbed HumanEval test set. Higher Pass@1 indicates better performance, while lower RD indicates greater robustness. Pass@1 is measured on the unperturbed and Pass@1\_perturbed on the perturbed datasets. }
    % \heng{Would be good to always place figures/tables at the page top}}
    \label{fig:RQ2}
    }
\end{figure*}

\subsubsection{Overall Trend}
To investigate the relationship between the proportion of perturbed data used during fine-tuning and resulting robustness, we trained models with proportions ranging from 0\% (unperturbed baseline) to 100\% (entirely perturbed). 
The results are summarized in Figure~\ref{fig:RQ2}, which reports average Pass@1 on both unperturbed and perturbed HumanEval test sets, as well as RD values.

Overall, perturbation-aware fine-tuning leads to substantial robustness improvements, but the \textbf{proportion of perturbed data plays a crucial role} in determining the trade-off between clean performance and robustness. 
Models trained with 50\%–70\% perturbed data achieve the best balance, maintaining strong Pass@1 on unperturbed test inputs while achieving lower RD on perturbed inputs. Their Pass@1 on unperturbed inputs drops by less than 5\%, while RD values are reduced by up to 20\% compared to the 0\% baseline. 
For example, at 70\% perturbation proportion, CodeLlama-7b-hf exhibits a relative RD reduction of 19.62\% compared to unperturbed fine-tuning, with only a marginal drop in clean Pass@1. 

Notably, CodeGen-2B exhibits a slightly different pattern: its RD curve remains relatively flat beyond 50\%, suggesting that this model is less sensitive to perturbation ratio compared to others. This indicates that perturbation proportion may play a model-dependent role in robustness gains.%\heng{not clear why text is italic here}

\subsubsection{Trade-Off and Non-Monotonic Trends}
At lower perturbation ratios ($<30\%$), the models receive insufficient exposure to distributional variation, leading to weaker robustness gains. 
At very high ratios ($\geq90\%$), robustness improves further, but at the cost of overall accuracy degradation.
% \heng{maybe ``overall'' is a better term than ``clean''?} 
This effect reflects a \textit{regularization–performance trade-off}, in which excessive exposure to perturbations forces the model to over-prioritize noisy input patterns, compromising task-specific precision.

Notably, we observe no strictly monotonic relationship between the perturbation ratio and robustness: performance varies with the random composition of perturbations across the sampled training sets. 
This variability aligns with our findings in RQ1, which show that not all perturbations 
% \Foutse{but which type of perturbation do we have in that figure 5? does this observation hold for any type of combinations of types of perturbations?} 
contribute equally to robustness. 
% In practice, the diversity of perturbations present in the training data is more critical than the absolute ratio of perturbed samples.\heng{I am not sure how this conclusion is supported: use numbers to support the comparison between diversity and perturbation ratio.}

\begin{findingbox}{RQ2}
% \textbf{Finding 1:} The proportion of perturbed data \armstrong{Provide concrete numbers of proportion here X\% }strongly affects the balance between robustness and clean performance. \heng{would be good to have numbers in the result discussion to support this, for example, what's the maximum variations of pass@1 and RD across different perturbation ratios.}\\
\textbf{Finding 2.1:}
% \Foutse{your numbering of findings is ambiguous, please fix it...people will not be able to refer to a specific finding in the paper if the numbering is repeated! may be use 2.1, 2.2}
% The proportion of perturbed data (ranging from 0\% to 100\%) substantially influences the balance between robustness and overall accuracy.  
% Across models, Pass@1 on unperturbed inputs decreases by less than 5\% on the perturbation ratio increases \Foutse{what do you mean by 'perturbation ratio increases'? please be precise! what is that ratio?}, while RD values are reduced 
% % \Foutse{reduced when? in which setting?} 
% by up to 20\% compared to the 0\% baseline\yang{(i.e. codellama-7b-hf at 70\% perturbatiion ratio)}.  
% This indicates that moderate perturbation exposure enhances robustness effectively without significant loss of accuracy, although the sensitivity to perturbation ratio remains model-dependent.\\
Finding 2.1: The proportion of perturbed data in the training mix strongly influences the trade-off between robustness and clean performance. Across models, Pass@1 on unperturbed inputs decreases by less than 5 percentage points as the perturbation ratio rises from 0\% to 100\%, while Relative Degradation (RD) on perturbed test sets is reduced by up to 20 percentage points compared to the 0\%-perturbation baseline (e.g., observed for CodeLlama-7B at a 70\% perturbation ratio). This suggests that moderate perturbation exposure can substantially improve robustness with only a minor impact on accuracy, although the optimal ratio remains model-dependent.

\textbf{Finding 2.2:}
Training with 50\%–70\% perturbed samples yields the most favorable robustness–accuracy trade-off.  
In this range, models achieve substantially lower RD values while maintaining high performance on the unperturbed test set, indicating that robustness can be improved without compromising standard accuracy.

% \textbf{Finding 2.2:} Training with 50\%–70\% perturbed samples provides the best trade-off, improving robustness while preserving clean \Foutse{what do you mean by 'clean' accuracy? Are you referring to the accuracy on dataset that hasn't been perturbed? please make sure to define that term because beyond the ML community your reviewer may not understand this!!!} accuracy.\\
% \textbf{Finding 3:} Diversity of perturbations is more influential than the raw proportion, as indicated by the non-monotonic performance trend.\heng{see comment above}
\end{findingbox}

\subsection{ \emph{\rqthree}}

\begin{figure*}[!ht]
{\small
    \centering
    \includegraphics[scale=0.65]{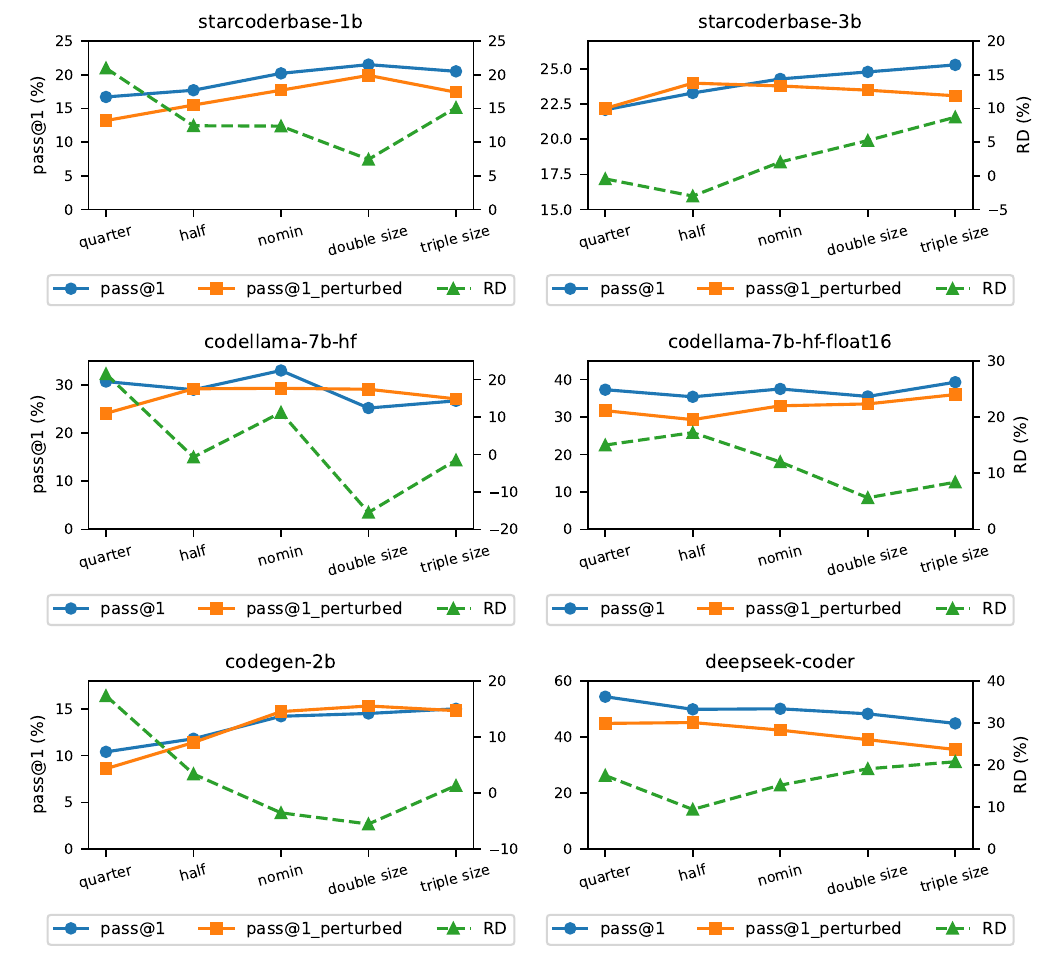}
   \caption{Performance and robustness of fine-tuned models with varying training dataset sizes, evaluated on the mix\_all\_level perturbed HumanEval test set. Higher Pass@1 indicates better performance, while lower RD indicates greater robustness. Pass@1 is measured on the unperturbed dataset, and Pass@1\_perturbed on the perturbed dataset. \textit{Nomin} denotes the original dataset size used in RQ1 and RQ2.}
    \label{fig:RQ3}
    }
\end{figure*}
 
\subsubsection{Moderate Scaling Improves Robustness}

We further examine how the \textbf{size of the training dataset} influences robustness. 
Models were fine-tuned on datasets ranging from one-quarter to three times the nominal size (30,087 samples), with a fixed 60\% perturbation ratio, which previously demonstrated the best robustness–accuracy trade-off. 
Figure~\ref{fig:RQ3} shows Pass@1 on both unperturbed and perturbed test sets, along with RD values.

Overall, \textbf{moderate dataset scaling improves robustness}, though the effect varies across models. 
For four out of six models (\texttt{StarCoderBase-1B}, \texttt{StarCoderBase-3B}, \texttt{CodeLlama-7B}, and \texttt{DeepSeek-Coder}), increasing the dataset size from original to double leads to substantial RD reductions (averaging 25–35\%) while maintaining stable Pass@1 on unperturbed data. 
For instance, \texttt{StarCoderBase-1B} trained on a double size of the dataset shows a 27.4\% RD reduction relative to the nominal size, with minimal change in clean accuracy.

In contrast, when the size of the dataset triples, \texttt{CodeGen-2B} exhibits an unusual trend: its RD decreases, indicating that larger training datasets up to triple size still provide additional robustness gains for this specific model. 
This model-dependent variation implies that dataset scaling interacts with model capacity and pre-training diversity in shaping robustness outcomes.

\subsubsection{Diminishing Returns and Noise Dilution}
Beyond moderate scaling, the benefits diminish. 
Expanding the dataset to 
% \Foutse{you mean beyond 3x? because in the previous section you just said that 3x yielded additional robustness!!! please double check to avoid contradicting yourself!}
triple size of the dataset does not yield further robustness improvements and sometimes even slightly reduces both Pass@1 and RD (observed in \texttt{CodeLlama-7B}, \texttt{StarCoderBase-3B}, and \texttt{DeepSeek-Coder}). 
This can be attributed to a \textit{signal-to-noise dilution effect}: as the dataset grows, the relative density of informative perturbations decreases, increasing the chance of redundant or less challenging examples that limit further robustness learning.

Conversely, reducing the dataset size (0.25× or 0.5×) results in lower Pass@1 and noticeably higher RD for all models. 
On average, these smaller datasets lead to a 15–25\% drop in clean accuracy and a 20–40\% increase in RD, reflecting insufficient exposure to diverse perturbation patterns.

Taken together, these results demonstrate that dataset size affects robustness in a nonlinear manner. 
Moderate expansion ensures exposure to a diverse but not redundant perturbation space, while excessive scaling dilutes the signal strength of informative examples.

\begin{findingbox}{RQ3}
\textbf{Finding 3.1:} 
% \Foutse{please adjust the numbering of your findings!!! 3.1, 3.2, 3.3}
Moderate dataset scaling (1×–2× original size, i.e., 30k–60k samples) yields the best robustness improvements, reducing RD by 25–35\% without degrading clean accuracy.\\
\textbf{Finding 3.2:} Very small datasets lack sufficient perturbation diversity, while excessively large datasets dilute informative noise, leading to diminishing returns.\\
\textbf{Finding 3.3:} The effect of dataset scaling is partially model-dependent, as models like \texttt{CodeGen-2B} continue to gain robustness even at larger scales.
\end{findingbox}

\subsection{Perturbation Categories}
We observe that the strongest robustness improvements resulted from the character-level, word-level, and mixed perturbations, which is to some extent supported by the findings of Zhu et al.~\cite{zhu2023promptrobust}. %In contrast, sentence-level perturbations provided narrower benefits. 
In contrast, sentence-level perturbations yielded more limited improvements. This finding highlights a significant implication when fine-tuning LLM4Code with perturbation categories (types), as we observe variability in their ability to regularize training. In particular, while a specific category of perturbations influences the LLM4code models to generalize beyond their training patterns, others lead to more localized robustness. %\heng{I don't see how this paragraph is useful. It's just a summary of results, not a discussion} 
% \Foutse{may be use a box environment for these actionable insight or something else that highlights them well!}\\
% \begin{Actionable_insight}
% Robustness-oriented fine-tuning should prioritize character-level, word-level and mixed perturbations, which closely mimic real-world input variability. On the other hand,  sentence-level perturbations are still valuable in contexts where typographical or lexical noise is frequent, such as novice programming environments.
% \end{Actionable_insight}
% \textbf{Actionable insight:} Robustness-oriented fine-tuning should prioritize character-level, word-level and mixed perturbations, which closely mimic real-world input variability. On the other hand,  sentence-level perturbations are still valuable in contexts where typographical or lexical noise is frequent, such as novice programming environments.

\subsection{Perturbation Ratio}
Varying the proportion of perturbed to unperturbed samples showed that robustness is maximized when 50–70\% of the dataset is perturbed. Beyond this range, we observed that improvements plateau, indicating that robustness is not simply proportional to the perturbation ratio; instead, the diversity of perturbation categories also has a greater impact, as noted by Ahsan et al.~\cite{ahsan2021effect}. This finding highlights that the robustness of the LLM4Code model is more visible when considering the balanced exposure to variability than simply maximizing the absolute share of perturbed data.

\textbf{Actionable insight:} Practitioners should avoid fine-tuning (training) LLM4Codes on extreme conditions, i.e., using only a few/limited perturbations, which would limit models' robustness; and also, using exclusively perturbed data would risk unnecessary performance degradation. A balanced ratio utilizing a mixture of approximately half perturbed and half unperturbed seems most effective. Moreover, sampling the perturbed datasets to increase diversity, rather than expanding them to cover a large portion of the training data, is a sensible and efficient choice when resources are limited.

%Moreover, it's commendable to sample the perturbed dataset to diversify it, rather than maximizing a large proportion of the training set, especially under constrained resource conditions. 

\subsection{Size of the Dataset }
Scaling the dataset revealed diminishing returns. Robustness improved when the dataset was doubled (from 30,087 to 60,174 samples), but further increasing it to three times its original size yielded minimal gains and in some cases even reduced performance. Prior work has also reported an inverse relationship between performance and excessive dataset size, as shown by Bahri et al.~\cite{bahri2024explaining} and Gordon et al.~\cite{gordon-etal-2021-data}. Our results suggest that robustness depends more on the quality and diversity of the perturbations than on the sheer volume of perturbed data, consistent with the findings of Howe et al.~\cite{howe2024exploring}.
%Scaling the dataset size revealed diminishing returns. Robustness improved when the dataset was increased to twice its nominal size (30,087–60,174 samples), but further enlargement to three times its original size provided little benefit and, in some cases, even reduced performance, prior works reported inverse proportiaonlity on perforrmance and dataset size, Bahri et al.~\cite{bahri2024explaining} and Gordon et al.~\cite{gordon-etal-2021-data}. 
%Based on our observations, its robustness turns out to be associated more with the quality and diversity of perturbations, rather than the volume of the unperturbed dataset, as noted by Howe et al.~\cite{howe2024exploring}.

\textbf{Actionable insight:} Moderate expansion (up to 2$\times$ the nominal size) is beneficial, but beyond this point the gains plateau. Developers should therefore focus on curating rich, diverse sets of perturbed data rather than disproportionately increasing the overall dataset size.

\section{Discussion}  \label{sec:discussion}

This study examines the impact of perturbation-aware fine-tuning on the robustness of LLM4Code, addressing three key research questions. Our findings demonstrate that when LLM4Code models are trained on perturbed datasets, they can significantly enhance resilience against adversarial inputs and also increase their reliability in real-world applications. However, we observe an overall performance degradation at the expense of the improvements of models' robustness; hence, a trade-off between the model's robustness and accuracy.  
% Furthermore, the variability observed across programming languages\heng{Is this discussed in the results} suggests that fine-tuning strategies may need to be adapted to programming language-specific characteristics. 
In the following subsections, we revisit each RQ, interpret the findings across RQs.
% and highlight actionable insights. 
We also complement robustness findings with an analysis of code quality and compliance.

% \heng{I suggest merge 6.1, 6.2, and 6.3 below into the RQ results in the previous section. Only keep across-RQ discussions and beyond-RQ discussions in this section.}\\
% \armstrong{$\uparrow$ Yes, I agree since we are reporting the observations.$\Leftarrow$}

\subsection{Lessons learnt across analyzing RQs}
Together, RQ1–RQ3 reveal that robustness is optimized by balancing quality, diversity, and moderate scaling of training data, rather than relying on extreme proportions or dataset sizes. Furthermore, this observation illustrates a transparent space for designing robustness-oriented fine-tuning:  
\begin{itemize}
    \item Use perturbation types that approximate realistic variability (RQ1).  
    % \item Maintain a reasonable balanced of perturbed-to-unperturbed ratio, thus, avoiding under- or over-generalization \Foutse{what do you mean by 'over generalization'?} (RQ2).  
    \item Maintain a balanced proportion of perturbed and unperturbed samples to avoid both under-generalization (i.e., the model fails to handle noisy inputs) and over-generalization (i.e., the model over-adapts to perturbed data and its performance on unperturbed inputs deteriorates) (RQ2).
    \item Expand the dataset %Datasets expansion 
    with moderation, while emphasizing diversity over size (RQ3).  
\end{itemize}

\subsection{Beyond the Robustness of LLM4Code: Generated Code Quality.}
% \armstrong{Simply report the statistic of sampling, how we got 380 and 383: What was the population, N, sample size, n, C.I., and C.L?}
% To complement robustness results, we evaluate the quality of code generated by LLM4Code from a sample of 380 from HumanEval and 383 from MBPP, \yang{corresponding to a 95\% confidence level for estimating mean quality scores with reasonable precision.\heng{explain why 380 and 383}}. 
To complement robustness results, we assess the quality of code generated by LLM4Code using randomly sampled subsets of 380 instances from HumanEval and 383 from MBPP. The sample sizes were determined using a standard calculator\footnote{https://www.calculator.net/sample-size-calculator.html} with a 95\% confidence level and a ±5\% margin of error, based on estimated population sizes of 31,878 (161 tasks × 198 samples) and 84,546 (427 × 198), respectively. These subsets ensure statistically reliable, yet practically manageable evaluations of average code quality.

\begin{figure}[ht]
\centering
\begin{tikzpicture}[scale=0.85, every node/.style={align=center,font=\small}]

% Define layer height
\def\h{2.0}

% Pyramid layers (top to bottom, wider as we go down)
\fill[red!80]       (0,6*\h) -- (-3,5*\h) -- (3,5*\h) -- cycle;
\fill[red!50]       (-3,5*\h) -- (3,5*\h) -- (4,4*\h) -- (-4,4*\h) -- cycle;
\fill[orange!70]    (-4,4*\h) -- (4,4*\h) -- (5,3*\h) -- (-5,3*\h) -- cycle;
\fill[yellow!70]    (-5,3*\h) -- (5,3*\h) -- (6,2*\h) -- (-6,2*\h) -- cycle;
\fill[green!50]     (-6,2*\h) -- (6,2*\h) -- (7,1*\h) -- (-7,1*\h) -- cycle;
\fill[gray!30]      (-7,1*\h) -- (7,1*\h) -- (8,0) -- (-8,0) -- cycle;

% Labels (aligned at layer midpoints)
\node at (0,5.3*\h) {\textbf{Fatal (F)} \\ Critical: \\ analysis cannot continue};
\node at (0,4.5*\h) {\textbf{Error (E)} \\ Tier 1: blockers (syntax-error, import-error) \\ Tier 2: API misuse (no-member, bad args)};
\node at (0,3.5*\h) {\textbf{Warning (W)} \\ Tier 1: high-risk (bare-except, broad-except) \\ Tier 2: moderate (shadowing, unused vars) \\ Tier 3: low (global stmt, reimport)};
\node at (0,2.5*\h) {\textbf{Refactor (R)} \\ High: complexity (too-many-branches) \\ Low: clarity (useless-option-value)};
\node at (0,1.5*\h) {\textbf{Convention (C)} \\ High: readability (invalid-name, missing-docstring) \\ Low: cosmetic (whitespace, newline)};
\node at (0,0.5*\h) {\textbf{Information (I)} \\ Informational only};

\end{tikzpicture}
\caption{Pylint severity taxonomy as a pyramid. 
Severity decreases from top (Fatal) to bottom (Information). 
Within categories like Error and Warning, editorial tiers highlight practical risk levels. %\heng{this figure is not needed as all the information is already in Table 12}
% \armstrong{we keep it.}
}
\label{fig:pylint-pyramid}
\end{figure}
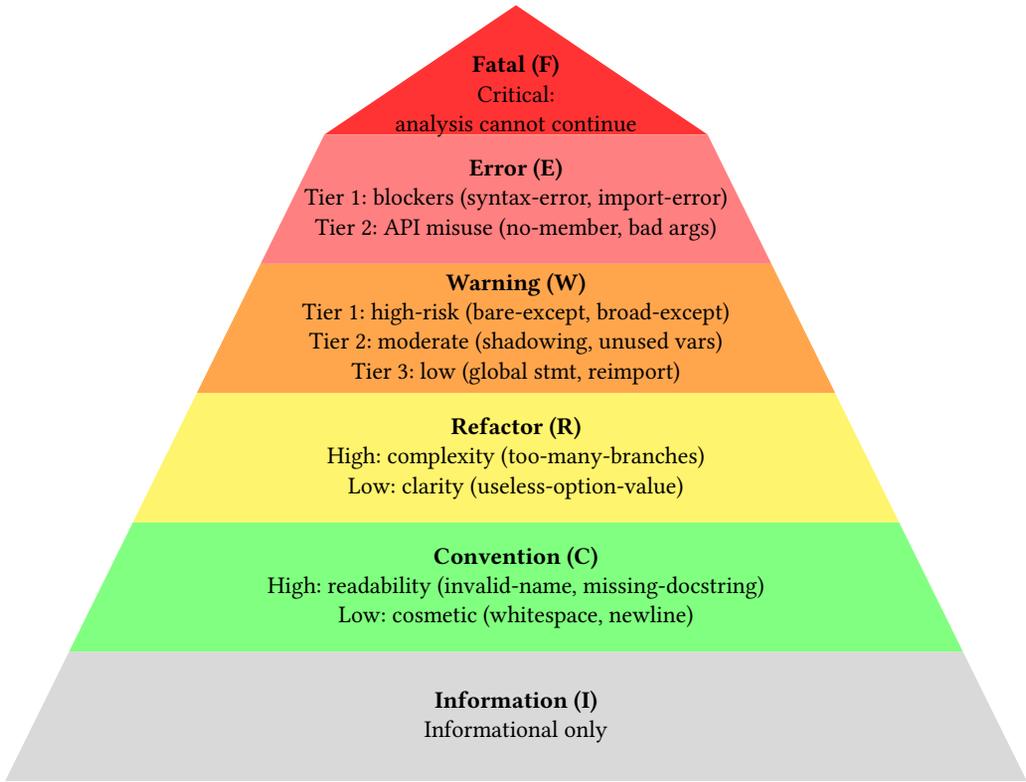

The Fine-tuned LLM4Code models consistently outperform the base models in style and syntactic correctness, with scores averaging 6.67/10 compared to 6.00/10. 
These gains were largely independent of perturbation type, ratio, or dataset size, suggesting that the act of fine-tuning itself is the primary driver of the improved quality in the generated code.
% \heng{Motivation: the purpose of this analysis}
We use Pylint\footnote{\url{https://pylint.readthedocs.io/en/stable/}} %as the tool to 
to assess the quality of the generated code. 
% \heng{the quality of the generated code, not ``generated...quality''} 
In Pylint, diagnostic messages are organized through a dual taxonomy of severity categories and rule groupings. As illustrated in Fig~\ref{fig:pylint-pyramid} and Table~\ref{tab:pylint-taxonomy-tiers}, severity follows a hierarchical structure from Fatal (F), where analysis cannot continue, through Error (E) and Warning (W), down to Refactor (R), Convention (C), and finally Information (I), which is purely descriptive. Errors and warnings are further stratified into editorial tiers that reflect practical risk levels, ranging from blockers, such as syntax errors (e.g., E0001), to moderate issues like variable shadowing (W0621), and low-level maintainability concerns, including redundant imports.

By contrast, the numeric component of the message code does not indicate severity but rather groups related rules. For example, the C01xx series covers naming conventions and documentation (e.g., C0103 for invalid naming, C0114 for missing module docstrings), C02xx focuses on Python idioms (e.g., C0200 recommending enumerate over range(len())), and C03xx addresses layout and formatting concerns (e.g., C0301 line length, C0303 trailing whitespace, C0305 redundant blank lines). These groupings serve to enhance readability and consistency but are not ranked in terms of risk. Thus, while the letter prefix establishes the severity hierarchy, the numeric suffix classifies the rule type, meaning that codes such as C0103 and C0114 are parallel checks within the same convention group rather than indicators of relative seriousness.

\small                                  % or \footnotesize
\setlength{\tabcolsep}{4pt}             % default ~6pt; tighten horizontally
\renewcommand{\arraystretch}{1.05} % default 1.2 with booktabs; tighten vertically a bit

\begin{table}[!ht]
\centering
\caption{Pylint message taxonomy with severity categories and suggested intra-category tiers. Severity tiers are editorial and indicate practical risk differences within each category. 
\textit{Category (F/E/W/R/C/I) = official Pylint classification. Exit Bit = how Pylint encodes this in exit codes (bitwise). 
Severity Tier = editorial overlay for distinguishing more vs.\ less severe rules inside the same category. 
Example: \texttt{W0702} (bare except) is Tier~1 because it can mask runtime crashes. 
\texttt{W0621} (redefined-outer-name) is Tier~2 because it is confusing but less catastrophic. 
Both are W (Warnings) in Pylint, but tiering makes their risk difference clear.}}
\label{tab:pylint-taxonomy-tiers}
\resizebox{\columnwidth}{!}{%
\begin{tabular}{l l l p{7.5cm}}
\toprule
\textbf{Category} & \textbf{Exit Bit} & \textbf{Severity Tier(s)} & \textbf{Description and Examples} \\
\midrule
\textbf{Fatal (F)} & 1  & Critical & Analysis cannot continue. Always stop. 
  Examples: \texttt{F0001 fatal}, \texttt{F0010 parse-error}. \\
\midrule
\textbf{Error (E)} & 2  & Tier 1 (blockers) & Code is invalid or will crash at runtime. 
  Examples: \texttt{E0001 syntax-error}, \texttt{E0401 import-error}, \texttt{E1120 no-value-for-parameter}. \\
\rowcolor{gray!10}
                   &    & Tier 2 (API misuse) & Likely logic bug, but analysis can continue. 
  Examples: \texttt{E1101 no-member}, \texttt{E1123 unexpected-keyword-arg}. \\
\midrule
\textbf{Warning (W)} & 4 & Tier 1 (high-risk) & Risky constructs that can hide runtime failures. 
  Examples: \texttt{W0702 bare-except}, \texttt{W0718 broad-exception-caught}, \texttt{W0102 dangerous-default-value}. \\
\rowcolor{gray!10}
                     &   & Tier 2 (moderate) & Likely logic mistakes / scoping issues. 
  Examples: \texttt{W0621 redefined-outer-name}, \texttt{W0612 unused-variable}. \\
                     &   & Tier 3 (low) & Maintainability odors / environment smells. 
  Examples: \texttt{W0603 global-statement}, \texttt{W0404 reimported-module}. \\
\midrule
\textbf{Refactor (R)} & 8 & High & Structural complexity that increases bug risk. 
  Examples: \texttt{R0912 too-many-branches}, \texttt{R0915 too-many-statements}. \\
\rowcolor{gray!10}
                       &   & Low & Clarity-only refactor suggestions. 
  Example: \texttt{R0022 useless-option-value}. \\
\midrule
\textbf{Convention (C)} & 16 & High & Style violations that affect readability/consistency. 
  Examples: \texttt{C0103 invalid-name}, \texttt{C0114 missing-module-docstring}. \\
\rowcolor{gray!10}
                         &    & Low & Cosmetic issues. 
  Examples: \texttt{C0303 trailing-whitespace}, \texttt{C0304 missing-final-newline}. \\
\midrule
\textbf{Information (I)} & -- & Info & Informational only, does not affect scoring. \\
\bottomrule
\end{tabular}}
\end{table}

\begin{table}[ht]
\centering
\caption{Warning Code Definitions for Code Compliance Analysis}
\begin{tabular}{c|l}
\hline
\textbf{Warning Code} & \textbf{Description} \\
\hline
E0001 & Critical error: Syntax error or invalid code \\
C0103 & Naming convention: Variable or function name does not conform to standards \\
C0114 & Missing module docstring: No docstring present at the top of the file \\
C0200 & Unnecessary list comprehension: Use of list comprehension with no effect \\
C0301 & Line too long: Exceeds maximum allowed characters per line \\
C0303 & Trailing whitespace: Presence of unnecessary whitespace at line end \\
C0305 & Trailing newlines: Excessive blank lines at the end of file \\
W0613 & Unused argument: Function argument not used in function body \\
W0621 & Variable name shadowing: Local variable name shadows a global variable \\
\hline
\end{tabular}
\label{tab:warning_codes}
\end{table}

\begin{table}[ht]
    \centering
    \caption{Code Compliance Warnings for Base model \texttt{codegen-2b} Models, and the training dataset is humaneval\_mix\_all\_level. FT w/o P means fine-tuned with the unperturbed dataset. Ft W/P means fine-tuned with perturbed dataset.}
    \begin{tabular}{l|c|c|c}
        \toprule
        \textbf{Warning Code} & \textbf{Base Model} & \textbf{Ft w/o P} & \textbf{Ft w/ P} \\
        \midrule
        C0114 & 12  & 22  & 4  \\
        C0103 & 12  & 32  & 2 \\
        C0305 & 11  & 0  & 0 \\
        C0200 & 10  & 0  & 0 \\
        E0001 & 8  & 0  & 0 \\
        C0301 & 2  & 2  & 0 \\
        C0303 & 1  & 11  & 2  \\
        W0613 & 1  & 20  & 4 \\
        \bottomrule
    \end{tabular}
    \label{tab:codegen-2b-warnings}
\end{table}

\begin{table}[ht]
    \centering
    \caption{Code Compliance Warnings for \texttt{codegen-2b} Models fintuned with different perturbed data ratio. $P_R$ means perturbed ratio.}
    \begin{tabular}{l|c|c|c|c|c}
        \toprule
        \textbf{Warning Code} & \textbf{$P_R$=10\%} & \textbf{$P_R$=30\%} & \textbf{$P_R$=50\%} & \textbf{$P_R$=70\%} & \textbf{$P_R$=90\%}\\
        \midrule
        C0303 & 5  & 5  & 5  & 5  & 5 \\
        C0114 & 10  & 10  & 10  & 10  & 10 \\
        C0103 & 14  & 15  & 13  & 14  & 15 \\
        W0621 & 8  & 10  & 6  & 8  & 10 \\
        \bottomrule
    \end{tabular}
    \label{tab:codegen-2b-ratio-warnings}
\end{table}

\begin{table}[ht]
    \centering
    \caption{Code Compliance Warnings for \texttt{codegen-2b} Models fintuned with different training data size.}
    \begin{tabular}{l|c|c|c|c}
        \toprule
        \textbf{Warning Code} & \textbf{data size=Quarter} & \textbf{data size=half} & \textbf{data size=double} & \textbf{data size=Triple}\\
        \midrule
        E0001 & 1  & 1  & 0 & 0 \\
        C0303 & 1  & 0  & 2 & 1 \\
        C0114 & 3  & 3  & 4 & 4 \\
        C0103 & 4  & 3  & 5 & 4 \\
        C0200 & 0  & 0  & 0 & 1 \\
        W0621 & 2  & 0  & 2 & 0 \\
        \bottomrule
    \end{tabular}
    \label{tab:codegen-2b-sample_size-warnings}
\end{table}

In addition, Tables~\ref{tab:warning_codes}–\ref{tab:codegen-2b-sample_size-warnings} highlight the importance of perturbation-aware fine-tuning, which shows a reduction in specific style-related warnings (e.g., missing docstrings), which is likely due to the improved models' generalization. Meanwhile, dataset scaling also shows a reduction in critical syntax errors (E0001). However, style and documentation issues (C0114, C0103) persistently indicate that fine-tuned models still do not fully internalize community coding standards. Half-precision training introduced slightly more stylistic warnings but did not generate new severe errors.

\textbf{Actionable insight:} Evaluating LLM4Code should extend beyond a one-dimention of functional correctness (Pass@1) to include robustness and maintainability. Fine-tuning reliably improves both robustness and readability; however, further interventions, such as style-aware fine-tuning~\cite{cai2024exploring} or multi-objective optimization~\cite{yang2024multi}, are necessary 
% \Foutse{how do you know that? did you tried it? if not, do you have a reference to back this claim?} 
to reduce documentation and stylistic violations.

\subsection{Synthesis}
It should be noted that perturbation-aware fine-tuning not only improves robustness but also the code quality, even though with noticeable trade-offs in the overall model performance. Our findings show that robustness is not solely determined by the quantity of data or the extent of perturbation. Instead, it hinges on several factors: selecting perturbation types that meaningfully reflect realistic input variability (as observed in RQ1), using balanced perturbed-to-unperturbed ratios that avoid both under- and over-adaptation (RQ2), and scaling the dataset size in moderation while prioritizing diversity over sheer volume (RQ3). 
These empirically grounded design choices collectively define how effectively robustness-oriented fine-tuning can be applied.
% , such as carefully selecting (empirical choice \Foutse{this is vague!! explain more clearly please!}) perturbation types, balanced ratios, and moderately scaled datasets. %Additionally, analyzing the quality of code generated by fine-tuned LLM4Code emphasizes that while robustness is important, it is not enough on its own; rather, we should also prioritize maintainability and adherence to code standards. These insights provide a roadmap for systematically improving LLM4Code in both research and practice.
Additionally, our analysis of the code generated by fine-tuned LLM4Code shows that robustness, while essential, is not sufficient on its own. Models must also produce code that is maintainable and aligned with established coding standards. These insights offer a roadmap for systematically improving LLM4Code in both research and practical applications.

\clearpage

\section{Implications} \label{sec:implication}
Our findings carry important implications for a range of stakeholders involved in the research, development, deployment, and governance of LLMs for code. 
Before discussing stakeholder-specific implications, we highlight a cross-cutting actionable insight derived from our results.

\begin{Actionable_insight}
\noindent
Robustness-oriented fine-tuning should prioritize character-level, word-level, and mixed perturbations, which closely mimic real-world input variability. Sentence-level perturbations are still valuable in contexts where typographical or lexical noise is frequent, such as novice programming environments.
\end{Actionable_insight}

\vspace{0.5em}
\noindent\textbf{For researchers.}  
This study shows how perturbation-based fine-tuning can serve as a systematic and replicable framework for analyzing robustness in code models. By varying perturbation types, dataset sizes, and perturbation ratios, researchers gain a controlled experimental setup to isolate robustness factors. This methodology can be extended to explore new perturbation types (e.g., code-specific noise, automated refactoring), additional programming languages, and cross-modal tasks that combine natural language with code. In doing so, future work can build a more comprehensive science of robustness in LLM4Code.

\textbf{For model developers.}  
Our results reveal a clear trade-off: modest reductions in Pass@1 accuracy can yield substantial gains in robustness. This emphasizes the need to balance first-choice accuracy with resilience under noisy or adversarial prompts. Developers of LLM4Code models can leverage perturbation-based fine-tuning to produce more dependable models, particularly in domains where robustness is critical (e.g., security-sensitive systems, embedded software). In addition, our analysis of dataset proportions and sizes provides concrete guidance for configuring fine-tuning strategies.

\textbf{For software developers and downstream users of AI coding assistants.}
Software engineers often submit prompts that vary in phrasing, completeness, or correctness, from typos to ambiguous descriptions. Robustness-oriented fine-tuned models are better equipped to handle such variability, reducing the risk of generating invalid, insecure, or misleading outputs. This can improve trust and reliability in practical workflows, especially in settings where users may not have the expertise to craft precise prompts.

\textbf{For platform providers and industry stakeholders.}  
Robustness should be considered a central dimension of safe deployment. Platforms integrating LLM4Code into IDEs, code review pipelines, or educational environments can use robustness-enhanced models to deliver safer and more reliable user experiences. For industry stakeholders, our findings highlight the importance of explicitly evaluating robustness, rather than relying solely on benchmark accuracy, before integrating LLM4Code models into production systems.

\textbf{For educators and learners.}  
As LLM4Code tools become embedded in programming education, robustness is critical to ensure that learners receive correct and reliable feedback. Perturbation-aware fine-tuned models can reduce the risk of exposing students to buggy, misleading, or non-standard code, thereby supporting effective pedagogy and skill development. Educators can also use robustness benchmarks as teaching resources to demonstrate the limits and safe use of AI coding assistants.

\textbf{For policymakers and regulators.}  
With increasing reliance on AI-generated code in safety-critical domains (e.g., healthcare, finance, infrastructure), regulators must ensure that deployed models are not only accurate but also robust to input variability and adversarial manipulation. Our framework offers a practical basis for developing robustness evaluation standards and certification procedures, which can help guide responsible adoption of LLM4Code systems in industry.

\textbf{For open-source communities.}  
Many state-of-the-art LLM4Code models are developed and maintained in open-source ecosystems. Our findings suggest that robustness benchmarks and perturbation-based fine-tuning strategies can be shared as community resources, allowing contributors to evaluate and improve models collaboratively. This can accelerate collective progress toward more reliable and sustainable open-source AI tools.

\section{Limitation And Future Works}\label{sec:limi}
While our study is comprehensive, it has limitations. Firstly, the number of perturbation types that we explored is limited and may not fully represent the variety of real-world code perturbations~\cite{mastropaolo2023robustness,jimenez2024swebench,liu2023your,huang2024effibench}. Secondly, because of the structure of the training dataset, we can only add perturbations on the task description part. We did not explore how perturbations on function names and variable names impact the robustness of fine-tuned models~\cite{du2024evaluating,le2023codechain}. Thirdly, because of the limited GPU resources, we use smaller models to replace large models. 
The results may differ for larger models with more than 8B parameters. 
%The results might be different from those larger models whose parameter size is larger than 8B. 
Also, because we use different fine-tuning methods for models with parameter size below and above 2B, we did not explored the impact of different fine-tuning methods on robustness.

% Additionally, although we did some tests on other programming languages such as Java and C++, the performance of fine-tuned models drop a lot in Java and C++.\heng{not in the results? Maybe the limitation can be phrased like: We focused our study on the ... language. Our findings may not generalize to other programming langauges. However, we did performe experiments consider other languages such as..... The results...} 
%We focused our study on the python language. Preliminary tests on Java and C++ showed notable performance degradation, indicating that our findings may not generalize across programming languages. 
%There are three possible reasons. Firstly, it might be the difference between Dynamical and static programming language. Secondly, it might be because the Java and C++ prompts lack fixed skeletal templates, making the models' implicit "code frame" highly senstive to minor changes. For small size models, they might not be able to recognize the Java and C++ frameworks in the sample, but large size models may be able to. Thirdly, it might be because the number of Java and C++ samples in training datasets is not enough. In our future work, we will solve drop performance in Java and C++.
We focused our study on Python. Preliminary tests on Java and C++ revealed notable performance degradation, suggesting that our findings may not fully generalize across programming languages.
Several factors may explain this gap. First, differences between dynamic (e.g., Python) and static (e.g., Java, C++) languages may affect how models reason about code. Second, the Java and C++ prompts lack fixed skeletal templates, making the models’ implicit “code frame” more sensitive to small variations; smaller models may fail to recognize these structural patterns, whereas larger models might. Third, the training datasets may contain fewer Java and C++ examples, limiting model familiarity with these languages.
Addressing the performance drop in Java and C++ will be a key direction for future work. 
%SafeCoder, which is our fine-tuning tool, focuses on generated more safe code. We only test the influence of the robustness on functionality, but we did not explore the impact on the security aspect. In future work, we will continue to explore the influence on the robustness of models on the security aspect.
SafeCoder, our fine-tuning tool, primarily targets the generation of safer code. In this work, we evaluated how robustness affects functionality, but we did not examine its impact on security. In future work, we plan to investigate how model robustness influences the security properties of the generated code. 
%Last but not least, 
% this\heng{what this means? be specific} approach 
%fine-tuning with perturbation training dataset usually improves the robustness of LLM4Code at the expense of slightly sacrificing model performance. We will explore the balance between the reducing performance and increasing robustness and also look for a better method to improve the robustness of LLM4Code.
Finally, fine-tuning with perturbed training data generally improves the robustness of LLM4Code, but often at the cost of a slight reduction in overall performance. In future work, we will explore how to better balance this trade-off and investigate alternative methods for improving robustness without compromising model performance.

\section{Threats to Validity} \label{sec:threats}

We discuss potential threats to the validity of our study following standard categories in empirical software engineering.

\textbf{Construct validity.}  
This concerns the suitability of the metrics and methods used to evaluate LLM4Code. While Pass@1 and Relative Degradation (RD) are widely adopted, they may not fully capture the complexity of code generation or the nuances of adversarial robustness. For example, automated unit tests can miss deeper semantic or security issues that manifest only at runtime or in specific deployment environments. To mitigate this, we complemented functional correctness with code quality assessments (\texttt{pylint}) to evaluate readability, style, and documentation compliance. Nonetheless, further development of richer evaluation frameworks is needed to capture the multifaceted nature of robustness in LLM4Code.

\textbf{Internal validity.}  
This refers to whether causal inferences drawn in the study are valid. A potential risk is that our perturbation techniques may not exhaustively represent all adversarial or noisy input scenarios. Moreover, the specific parameters used to generate perturbations (e.g., substitution rates, translation quality) may influence results. To mitigate this, we adopted multiple perturbation types across three levels (character, word, sentence), evaluated perturbation datasets independently by two authors, and resolved disagreements collaboratively. Still, unseen adversarial strategies may produce different outcomes.

\textbf{External validity.}  
This relates to the generalizability of our findings. We relied on HumanEval and MBPP benchmarks, which, although widely used in prior work, do not fully represent the diversity of real-world software engineering tasks. Industrial code often contains multi-module dependencies, build configurations, or domain-specific APIs that are not reflected in these benchmarks. Future work should expand evaluation to larger and more diverse datasets (e.g., CodeNet, BigCode) and include industry case studies to increase external validity.

\textbf{Conclusion validity.}  
This refers to the degree to which our conclusions are statistically reliable. While we applied statistical tests such as the Wilcoxon signed-rank test with effect sizes, the number of models and samples per experiment is limited, and replication with larger-scale experiments may yield additional insights. We mitigated random variation by averaging results over multiple runs with different seeds, but stochastic effects in training cannot be completely ruled out. Thus, while our conclusions are statistically significant, they should be interpreted with caution until replicated by independent studies. %\Foutse{what about family-wise errors stemming from multiple comparisons? did you applied p-value corrections? for example using Bonferroni or Holm-Bonferroni approaches?}

\textbf{Reliability.}  
This concerns the replicability of our study. We have released our perturbation datasets, fine-tuning configurations, and evaluation scripts in a replication package to ensure transparency and reproducibility~\cite{ReplicationPackage} 
% \Foutse{please add the reference to that package here!}. 
However, differences in hardware, software environments, or model checkpoints may cause slight variations in results. Ensuring long-term accessibility of replication artifacts is essential to support future verification and extension of our findings. We will release our replication package on zenodo on the final revision.
% \Foutse{true but how did we achieve that? did you upload on zenodo for example? since it is perpetual there!}

\section{Conclusions} \label{sec:conclusions}
In conclusion, this research systematically investigates the robustness of Large Language Models for Code (LLM4Code) under various perturbations or adversarial attacks. We created comprehensive perturbed datasets to assess the impact of perturbations at character, word, and sentence levels. Experimental results indicate that fine-tuning LLM4Code models with perturbed datasets enhances their robustness, albeit typically accompanied by a slight decline in performance. 
% Additionally, the robustness varies significantly across different programming languages\heng{not clear how this is supported in the results}, 
% \armstrong{Simple state the languages we used in the experiment} underscoring the complexity of fine-tuning multilingual code models. 
This study provides empirical insights into the relationship between robustness and perturbation characteristics, perturbation ratios, and training set size, offering practical considerations for deploying robust LLM4Code models in realistic software development scenarios.

%%
%% The acknowledgments section is defined using the "acks" environment
%% (and NOT an unnumbered section). This ensures the proper
%% identification of the section in the article metadata, and the
%% consistent spelling of the heading.

\noindent\textbf{Data Availability Statement} The replication package, including our datasets and more results, is available publicly at: 
\url{https://tinyurl.com/57rkutkd}
\begin{acks}
This work was supported by: Fonds de Recherche du Québec (FRQ), the Canadian Institute for Advanced Research (CIFAR) as well as the DEEL project CRDPJ 537462-18 funded by the Natural Sciences and Engineering Research Council of Canada (NSERC) and the Consortium for Research and Innovation in Aerospace in Québec (CRIAQ), together with its industrial partners Thales Canada inc, Bell Textron Canada Limited, CAE inc and Bombardier inc.
\end{acks}

%%
%% The next two lines define the bibliography style to be used, and
%% the bibliography file.
%\bibliographystyle{ACM-Reference-Format}
\bibliographystyle{IEEEtran}
\bibliography{references,software}

@String{Academic = "Academic Press" }

@String{Springer = "Springer-Verlag" }

@article{chen2021evaluating,
  title={Evaluating large language models trained on code},
  author={Chen, Mark and Tworek, Jerry and Jun, Heewoo and Yuan, Qiming and Pinto, Henrique Ponde de Oliveira and Kaplan, Jared and Edwards, Harri and Burda, Yuri and Joseph, Nicholas and Brockman, Greg and others},
  journal={arXiv preprint arXiv:2107.03374},
  year={2021}
}

@article{austin2021program,
  title={Program synthesis with large language models},
  author={Austin, Jacob and Odena, Augustus and Nye, Maxwell and Bosma, Maarten and Michalewski, Henryk and Dohan, David and Jiang, Ellen and Cai, Carrie and Terry, Michael and Le, Quoc and others},
  journal={arXiv preprint arXiv:2108.07732},
  year={2021}
}

@article{yuan2019adversarial,
  title={Adversarial examples: Attacks and defenses for deep learning},
  author={Yuan, Xiaoyong and He, Pan and Zhu, Qile and Li, Xiaolin},
  journal={IEEE transactions on neural networks and learning systems},
  volume={30},
  number={9},
  pages={2805--2824},
  year={2019},
  publisher={IEEE}
}

@article{he2024instruction,
  title={Instruction tuning for secure code generation},
  author={He, Jingxuan and Vero, Mark and Krasnopolska, Gabriela and Vechev, Martin},
  journal={arXiv preprint arXiv:2402.09497},
  year={2024}
}

@article{li2023starcoder,
  title={Starcoder: may the source be with you!},
  author={Li, Raymond and Allal, Loubna Ben and Zi, Yangtian and Muennighoff, Niklas and Kocetkov, Denis and Mou, Chenghao and Marone, Marc and Akiki, Christopher and Li, Jia and Chim, Jenny and others},
  journal={arXiv preprint arXiv:2305.06161},
  year={2023}
}

@article{guo2024deepseek,
  title={DeepSeek-Coder: When the Large Language Model Meets Programming--The Rise of Code Intelligence},
  author={Guo, Daya and Zhu, Qihao and Yang, Dejian and Xie, Zhenda and Dong, Kai and Zhang, Wentao and Chen, Guanting and Bi, Xiao and Wu, Yu and Li, YK and others},
  journal={arXiv preprint arXiv:2401.14196},
  year={2024}
}

@article{nijkamp2022codegen,
  title={Codegen: An open large language model for code with multi-turn program synthesis},
  author={Nijkamp, Erik and Pang, Bo and Hayashi, Hiroaki and Tu, Lifu and Wang, Huan and Zhou, Yingbo and Savarese, Silvio and Xiong, Caiming},
  journal={arXiv preprint arXiv:2203.13474},
  year={2022}
}

@article{roziere2023code,
  title={Code llama: Open foundation models for code},
  author={Roziere, Baptiste and Gehring, Jonas and Gloeckle, Fabian and Sootla, Sten and Gat, Itai and Tan, Xiaoqing Ellen and Adi, Yossi and Liu, Jingyu and Remez, Tal and Rapin, J{\'e}r{\'e}my and others},
  journal={arXiv preprint arXiv:2308.12950},
  year={2023}
}

@inproceedings{zheng2023codegeex,
  title={Codegeex: A pre-trained model for code generation with multilingual benchmarking on humaneval-x},
  author={Zheng, Qinkai and Xia, Xiao and Zou, Xu and Dong, Yuxiao and Wang, Shan and Xue, Yufei and Shen, Lei and Wang, Zihan and Wang, Andi and Li, Yang and others},
  booktitle={Proceedings of the 29th ACM SIGKDD Conference on Knowledge Discovery and Data Mining},
  pages={5673--5684},
  year={2023}
}

@article{fried2022incoder,
  title={Incoder: A generative model for code infilling and synthesis},
  author={Fried, Daniel and Aghajanyan, Armen and Lin, Jessy and Wang, Sida and Wallace, Eric and Shi, Freda and Zhong, Ruiqi and Yih, Wen-tau and Zettlemoyer, Luke and Lewis, Mike},
  journal={arXiv preprint arXiv:2204.05999},
  year={2022}
}

@inproceedings{batouta2016automation,
  title={Automation in code generation: Tertiary and systematic mapping review},
  author={Batouta, Zouhair Ibn and Dehbi, Rachid and Talea, Mohammed and Hajoui, Omar},
  booktitle={2016 4th IEEE International Colloquium on Information Science and Technology (CiSt)},
  pages={200--205},
  year={2016},
  organization={IEEE}
}

@inproceedings{bielik2020adversarial,
  title={Adversarial robustness for code},
  author={Bielik, Pavol and Vechev, Martin},
  booktitle={International Conference on Machine Learning},
  pages={896--907},
  year={2020},
  organization={PMLR}
}

@inproceedings{carlini2017towards,
  title={Towards evaluating the robustness of neural networks},
  author={Carlini, Nicholas and Wagner, David},
  booktitle={2017 ieee symposium on security and privacy (sp)},
  pages={39--57},
  year={2017},
  organization={Ieee}
}

@article{metzen2017detecting,
  title={On detecting adversarial perturbations},
  author={Metzen, Jan Hendrik and Genewein, Tim and Fischer, Volker and Bischoff, Bastian},
  journal={arXiv preprint arXiv:1702.04267},
  year={2017}
}

@article{pearce2025asleep,
  title={Asleep at the keyboard? assessing the security of github copilot’s code contributions},
  author={Pearce, Hammond and Ahmad, Baleegh and Tan, Benjamin and Dolan-Gavitt, Brendan and Karri, Ramesh},
  journal={Communications of the ACM},
  volume={68},
  number={2},
  pages={96--105},
  year={2025},
  publisher={ACM New York, NY, USA}
}

@inproceedings{jha2023codeattack,
  title={Codeattack: Code-based adversarial attacks for pre-trained programming language models},
  author={Jha, Akshita and Reddy, Chandan K},
  booktitle={Proceedings of the AAAI Conference on Artificial Intelligence},
  volume={37},
  pages={14892--14900},
  year={2023}
}

@inproceedings{mastropaolo2023robustness,
  title={On the robustness of code generation techniques: An empirical study on github copilot},
  author={Mastropaolo, Antonio and Pascarella, Luca and Guglielmi, Emanuela and Ciniselli, Matteo and Scalabrino, Simone and Oliveto, Rocco and Bavota, Gabriele},
  booktitle={2023 IEEE/ACM 45th International Conference on Software Engineering (ICSE)},
  pages={2149--2160},
  year={2023},
  organization={IEEE}
}

@article{improta2025enhancing,
  title={Enhancing robustness of ai offensive code generators via data augmentation},
  author={Improta, Cristina and Liguori, Pietro and Natella, Roberto and Cukic, Bojan and Cotroneo, Domenico},
  journal={Empirical Software Engineering},
  volume={30},
  number={1},
  pages={7},
  year={2025},
  publisher={Springer}
}

@article{weyssow2023exploring,
  title={Exploring parameter-efficient fine-tuning techniques for code generation with large language models},
  author={Weyssow, Martin and Zhou, Xin and Kim, Kisub and Lo, David and Sahraoui, Houari},
  journal={ACM Transactions on Software Engineering and Methodology},
  year={2023},
  publisher={ACM New York, NY}
}

@inproceedings{li2024fine,
  title={Fine tuning large language model for secure code generation},
  author={Li, Junjie and Sangalay, Aseem and Cheng, Cheng and Tian, Yuan and Yang, Jinqiu},
  booktitle={Proceedings of the 2024 IEEE/ACM First International Conference on AI Foundation Models and Software Engineering},
  pages={86--90},
  year={2024}
}

@inproceedings{shi2023towards,
  title={Towards efficient fine-tuning of pre-trained code models: An experimental study and beyond},
  author={Shi, Ensheng and Wang, Yanlin and Zhang, Hongyu and Du, Lun and Han, Shi and Zhang, Dongmei and Sun, Hongbin},
  booktitle={Proceedings of the 32nd ACM SIGSOFT International Symposium on Software Testing and Analysis},
  pages={39--51},
  year={2023}
}

@article{guo2020graphcodebert,
  title={Graphcodebert: Pre-training code representations with data flow},
  author={Guo, Daya and Ren, Shuo and Lu, Shuai and Feng, Zhangyin and Tang, Duyu and Liu, Shujie and Zhou, Long and Duan, Nan and Svyatkovskiy, Alexey and Fu, Shengyu and others},
  journal={arXiv preprint arXiv:2009.08366},
  year={2020}
}

@article{feng2020codebert,
  title={Codebert: A pre-trained model for programming and natural languages},
  author={Feng, Zhangyin and Guo, Daya and Tang, Duyu and Duan, Nan and Feng, Xiaocheng and Gong, Ming and Shou, Linjun and Qin, Bing and Liu, Ting and Jiang, Daxin and others},
  journal={arXiv preprint arXiv:2002.08155},
  year={2020}
}

@article{alon2019code2vec,
  title={code2vec: Learning distributed representations of code},
  author={Alon, Uri and Zilberstein, Meital and Levy, Omer and Yahav, Eran},
  journal={Proceedings of the ACM on Programming Languages},
  volume={3},
  number={POPL},
  pages={1--29},
  year={2019},
  publisher={ACM New York, NY, USA}
}

@article{alon2018code2seq,
  title={code2seq: Generating sequences from structured representations of code},
  author={Alon, Uri and Brody, Shaked and Levy, Omer and Yahav, Eran},
  journal={arXiv preprint arXiv:1808.01400},
  year={2018}
}

@article{wang2022recode,
  title={ReCode: Robustness Evaluation of Code Generation Models},
  author={Wang, Shiqi and Li, Zheng and Qian, Haifeng and Yang, Chenghao and Wang, Zijian and Shang, Mingyue and Kumar, Varun and Tan, Samson and Ray, Baishakhi and Bhatia, Parminder and others},
  journal={arXiv preprint arXiv:2212.10264},
  year={2022}
}

@article{liu2025adversarialattackclassificationrobustness,
author = {Liu, Yang and Foundjem, Armstrong and Khomh, Foutse and Li, Heng},
title = {Adversarial attack classification and robustness testing for large language models for code},
year = {2025},
issue_date = {May 2025},
publisher = {Kluwer Academic Publishers},
address = {USA},
volume = {30},
number = {5},
issn = {1382-3256},
url = {https://doi.org/10.1007/s10664-025-10693-3},
doi = {10.1007/s10664-025-10693-3},
journal = {Empirical Softw. Engg.},
month = aug,
numpages = {57}
}

@article{thakur2024autotrain,
  title={AutoTrain: No-code training for state-of-the-art models},
  author={Thakur, Abhishek},
  journal={arXiv preprint arXiv:2410.15735},
  year={2024}
}

@article{lu2021codexglue,
  title={Codexglue: A machine learning benchmark dataset for code understanding and generation},
  author={Lu, Shuai and Guo, Daya and Ren, Shuo and Huang, Junjie and Svyatkovskiy, Alexey and Blanco, Ambrosio and Clement, Colin and Drain, Dawn and Jiang, Daxin and Tang, Duyu and others},
  journal={arXiv preprint arXiv:2102.04664},
  year={2021}
}

@article{allal2023santacoder,
  title={Santacoder: don't reach for the stars!},
  author={Allal, Loubna Ben and Li, Raymond and Kocetkov, Denis and Mou, Chenghao and Akiki, Christopher and Ferrandis, Carlos Munoz and Muennighoff, Niklas and Mishra, Mayank and Gu, Alex and Dey, Manan and others},
  journal={arXiv preprint arXiv:2301.03988},
  year={2023}
}

@article{reimers2017reporting,
  title={Reporting score distributions makes a difference: Performance study of LSTM-networks for sequence tagging},
  author={Reimers, Nils and Gurevych, Iryna},
  journal={arXiv preprint arXiv:1707.09861},
  year={2017}
}

@article{kingma2015adam,
  title={Adam: A method for stochastic optimization},
  author={Kingma, Diederik P and Ba, Jimmy},
  journal={International Conference on Learning Representations (ICLR)},
  year={2015},
  url={https://arxiv.org/abs/1412.6980}
}

@inproceedings{pascanu2013difficulty,
  title={On the difficulty of training recurrent neural networks},
  author={Pascanu, Razvan and Mikolov, Tomas and Bengio, Yoshua},
  booktitle={International Conference on Machine Learning (ICML)},
  pages={1310--1318},
  year={2013}
}

@article{hu2022lora,
  title={LoRA: Low-Rank Adaptation of Large Language Models},
  author={Hu, Edward J and Shen, Yelong and Wallis, Phillip and Allen-Zhu, Zeyuan and Li, Yuanzhi and Wang, Lu and Wang, Weizhu},
  journal={International Conference on Learning Representations (ICLR)},
  year={2022},
  url={https://arxiv.org/abs/2106.09685}
}

@misc{ReplicationPackage,
      title={Improving the Robustness of Large Language Models in Code Tasks via Fine-tuning with Perturbed Data}, 
      author={Anonymous},
      year={2025},
      url={https://tinyurl.com/57rkutkd}, 
}

@inproceedings{zhu2023promptrobust,
  title={Promptrobust: Towards evaluating the robustness of large language models on adversarial prompts},
  author={Zhu, Kaijie and Wang, Jindong and Zhou, Jiaheng and Wang, Zichen and Chen, Hao and Wang, Yidong and Yang, Linyi and Ye, Wei and Zhang, Yue and Gong, Neil and others},
  booktitle={Proceedings of the 1st ACM workshop on large AI systems and models with privacy and safety analysis},
  pages={57--68},
  year={2023}
}

@article{ahsan2021effect,
  title={Effect of data scaling methods on machine learning algorithms and model performance},
  author={Ahsan, Md Manjurul and Mahmud, MA Parvez and Saha, Pritom Kumar and Gupta, Kishor Datta and Siddique, Zahed},
  journal={Technologies},
  volume={9},
  number={3},
  pages={52},
  year={2021},
  publisher={MDPI}
}

@article{bahri2024explaining,
  title={Explaining neural scaling laws},
  author={Bahri, Yasaman and Dyer, Ethan and Kaplan, Jared and Lee, Jaehoon and Sharma, Utkarsh},
  journal={Proceedings of the National Academy of Sciences},
  volume={121},
  number={27},
  pages={e2311878121},
  year={2024},
  publisher={National Academy of Sciences}
}

@inproceedings{gordon-etal-2021-data,
    title = "Data and Parameter Scaling Laws for Neural Machine Translation",
    author = "Gordon, Mitchell A  and
      Duh, Kevin  and
      Kaplan, Jared",
    editor = "Moens, Marie-Francine  and
      Huang, Xuanjing  and
      Specia, Lucia  and
      Yih, Scott Wen-tau",
    booktitle = "Proceedings of the 2021 Conference on Empirical Methods in Natural Language Processing",
    month = nov,
    year = "2021",
    address = "Online and Punta Cana, Dominican Republic",
    publisher = "Association for Computational Linguistics",
    url = "https://aclanthology.org/2021.emnlp-main.478/",
    doi = "10.18653/v1/2021.emnlp-main.478",
    pages = "5915--5922"
}

@inproceedings{howe2024exploring,
  title={Exploring scaling trends in llm robustness},
  author={Howe, Nikolaus HR and Zaj{\k{a}}c, Micha{\l} and McKenzie, Ian R and Hollinsworth, Oskar John and Bacon, Pierre-Luc and Gleave, Adam},
  booktitle={ICML 2024 Next Generation of AI Safety Workshop},
  year={2024}
}

@inproceedings{ribeiro-etal-2020-beyond,
    title = "Beyond Accuracy: Behavioral Testing of {NLP} Models with {C}heck{L}ist",
    author = "Ribeiro, Marco Tulio  and
      Wu, Tongshuang  and
      Guestrin, Carlos  and
      Singh, Sameer",
    editor = "Jurafsky, Dan  and
      Chai, Joyce  and
      Schluter, Natalie  and
      Tetreault, Joel",
    booktitle = "Proceedings of the 58th Annual Meeting of the Association for Computational Linguistics",
    month = jul,
    year = "2020",
    address = "Online",
    publisher = "Association for Computational Linguistics",
    url = "https://aclanthology.org/2020.acl-main.442/",
    doi = "10.18653/v1/2020.acl-main.442",
    pages = "4902--4912"
}

@inproceedings{morris-etal-2020-textattack,
    title = "{T}ext{A}ttack: A Framework for Adversarial Attacks, Data Augmentation, and Adversarial Training in {NLP}",
    author = "Morris, John  and
      Lifland, Eli  and
      Yoo, Jin Yong  and
      Grigsby, Jake  and
      Jin, Di  and
      Qi, Yanjun",
    editor = "Liu, Qun  and
      Schlangen, David",
    booktitle = "Proceedings of the 2020 Conference on Empirical Methods in Natural Language Processing: System Demonstrations",
    month = oct,
    year = "2020",
    address = "Online",
    publisher = "Association for Computational Linguistics",
    url = "https://aclanthology.org/2020.emnlp-demos.16/",
    doi = "10.18653/v1/2020.emnlp-demos.16",
    pages = "119--126"
}

@inproceedings{jin2020bert,
  title={Is bert really robust? a strong baseline for natural language attack on text classification and entailment},
  author={Jin, Di and Jin, Zhijing and Zhou, Joey Tianyi and Szolovits, Peter},
  booktitle={Proceedings of the AAAI conference on artificial intelligence},
  volume={34},
  number={05},
  pages={8018--8025},
  year={2020}
}

@inproceedings{wang-etal-2021-textflint,
    title = "{T}ext{F}lint: Unified Multilingual Robustness Evaluation Toolkit for Natural Language Processing",
    author = "Wang, Xiao  and
      Liu, Qin  and
      Gui, Tao  and
      Zhang, Qi  and
      Zou, Yicheng  and
      Zhou, Xin  and
      Ye, Jiacheng  and
      Zhang, Yongxin  and
      Zheng, Rui  and
      Pang, Zexiong  and
      Wu, Qinzhuo  and
      Li, Zhengyan  and
      Zhang, Chong  and
      Ma, Ruotian  and
      Fei, Zichu  and
      Cai, Ruijian  and
      Zhao, Jun  and
      Hu, Xingwu  and
      Yan, Zhiheng  and
      Tan, Yiding  and
      Hu, Yuan  and
      Bian, Qiyuan  and
      Liu, Zhihua  and
      Qin, Shan  and
      Zhu, Bolin  and
      Xing, Xiaoyu  and
      Fu, Jinlan  and
      Zhang, Yue  and
      Peng, Minlong  and
      Zheng, Xiaoqing  and
      Zhou, Yaqian  and
      Wei, Zhongyu  and
      Qiu, Xipeng  and
      Huang, Xuanjing",
    editor = "Ji, Heng  and
      Park, Jong C.  and
      Xia, Rui",
    booktitle = "Proceedings of the 59th Annual Meeting of the Association for Computational Linguistics and the 11th International Joint Conference on Natural Language Processing: System Demonstrations",
    month = aug,
    year = "2021",
    address = "Online",
    publisher = "Association for Computational Linguistics",
    url = "https://aclanthology.org/2021.acl-demo.41/",
    doi = "10.18653/v1/2021.acl-demo.41",
    pages = "347--355"
}

@inproceedings{hendrycks2021measuring,
 author = {Hendrycks, Dan and Basart, Steven and Kadavath, Saurav and Mazeika, Mantas and Arora, Akul and Guo, Ethan and Burns, Collin and Puranik, Samir and He, Horace and Song, Dawn and Steinhardt, Jacob},
 booktitle = {Proceedings of the Neural Information Processing Systems Track on Datasets and Benchmarks},
 editor = {J. Vanschoren and S. Yeung},
 pages = {},
 title = {Measuring Coding Challenge Competence With APPS},
 url = {https://datasets-benchmarks-proceedings.neurips.cc/paper_files/paper/2021/file/c24cd76e1ce41366a4bbe8a49b02a028-Paper-round2.pdf},
 volume = {1},
 year = {2021}
}

@inproceedings{jimenez2024swebench,
title={{SWE}-bench: Can Language Models Resolve Real-world Github Issues?},
author={Carlos E Jimenez and John Yang and Alexander Wettig and Shunyu Yao and Kexin Pei and Ofir Press and Karthik R Narasimhan},
booktitle={The Twelfth International Conference on Learning Representations},
year={2024},
url={https://openreview.net/forum?id=VTF8yNQM66}
}

@article{dettmers2023qlora,
  title={Qlora: Efficient finetuning of quantized llms},
  author={Dettmers, Tim and Pagnoni, Artidoro and Holtzman, Ari and Zettlemoyer, Luke},
  journal={Advances in neural information processing systems},
  volume={36},
  pages={10088--10115},
  year={2023}
}

@inproceedings{dror2018hitchhiker,
  title={The hitchhiker’s guide to testing statistical significance in natural language processing},
  author={Dror, Rotem and Baumer, Gili and Shlomov, Segev and Reichart, Roi},
  booktitle={Proceedings of the 56th Annual Meeting of the Association for Computational Linguistics (Volume 1: Long Papers)},
  pages={1383--1392},
  year={2018}
}

@inproceedings{du2024evaluating,
  title={Evaluating large language models in class-level code generation},
  author={Du, Xueying and Liu, Mingwei and Wang, Kaixin and Wang, Hanlin and Liu, Junwei and Chen, Yixuan and Feng, Jiayi and Sha, Chaofeng and Peng, Xin and Lou, Yiling},
  booktitle={Proceedings of the IEEE/ACM 46th International Conference on Software Engineering},
  pages={1--13},
  year={2024}
}

@article{hou2024large,
  title={Large language models for software engineering: A systematic literature review},
  author={Hou, Xinyi and Zhao, Yanjie and Liu, Yue and Yang, Zhou and Wang, Kailong and Li, Li and Luo, Xiapu and Lo, David and Grundy, John and Wang, Haoyu},
  journal={ACM Transactions on Software Engineering and Methodology},
  volume={33},
  number={8},
  pages={1--79},
  year={2024},
  publisher={ACM New York, NY}
}

@article{wang2024software,
  title={Software testing with large language models: Survey, landscape, and vision},
  author={Wang, Junjie and Huang, Yuchao and Chen, Chunyang and Liu, Zhe and Wang, Song and Wang, Qing},
  journal={IEEE Transactions on Software Engineering},
  volume={50},
  number={4},
  pages={911--936},
  year={2024},
  publisher={IEEE}
}

@article{liu2023your,
  title={Is your code generated by chatgpt really correct? rigorous evaluation of large language models for code generation},
  author={Liu, Jiawei and Xia, Chunqiu Steven and Wang, Yuyao and Zhang, Lingming},
  journal={Advances in Neural Information Processing Systems},
  volume={36},
  pages={21558--21572},
  year={2023}
}

@article{le2023codechain,
  title={Codechain: Towards modular code generation through chain of self-revisions with representative sub-modules},
  author={Le, Hung and Chen, Hailin and Saha, Amrita and Gokul, Akash and Sahoo, Doyen and Joty, Shafiq},
  journal={arXiv preprint arXiv:2310.08992},
  year={2023}
}

@article{huang2024effibench,
  title={Effibench: Benchmarking the efficiency of automatically generated code},
  author={Huang, Dong and Qing, Yuhao and Shang, Weiyi and Cui, Heming and Zhang, Jie M},
  journal={Advances in Neural Information Processing Systems},
  volume={37},
  pages={11506--11544},
  year={2024}
}

@inproceedings{sidiropoulos2024improving,
  title={Improving the robustness of dense retrievers against typos via multi-positive contrastive learning},
  author={Sidiropoulos, Georgios and Kanoulas, Evangelos},
  booktitle={European Conference on Information Retrieval},
  pages={297--305},
  year={2024},
  organization={Springer}
}

@article{tang2023data,
  title={Data augmentation methods for enhancing robustness in text classification tasks},
  author={Tang, Huidong and Kamei, Sayaka and Morimoto, Yasuhiko},
  journal={Algorithms},
  volume={16},
  number={1},
  pages={59},
  year={2023},
  publisher={MDPI}
}

@article{adam2014method,
  title={A method for stochastic optimization},
  author={Adam, Kingma DP Ba J and others},
  journal={arXiv preprint arXiv:1412.6980},
  volume={1412},
  number={6},
  year={2014}
}

@article{kocetkov2022stack,
  title={The stack: 3 tb of permissively licensed source code},
  author={Kocetkov, Denis and Li, Raymond and Allal, Loubna Ben and Li, Jia and Mou, Chenghao and Ferrandis, Carlos Mu{\~n}oz and Jernite, Yacine and Mitchell, Margaret and Hughes, Sean and Wolf, Thomas and others},
  journal={arXiv preprint arXiv:2211.15533},
  year={2022}
}

@inproceedings{chen2022adversarial,
  title={Adversarial training for improving model robustness? Look at both prediction and interpretation},
  author={Chen, Hanjie and Ji, Yangfeng},
  booktitle={Proceedings of the AAAI conference on artificial intelligence},
  volume={36},
  number={10},
  pages={10463--10472},
  year={2022}
}

@article{cai2024exploring,
  title={Exploring Style-Robust Scene Text Detection via Style-Aware Learning},
  author={Cai, Yuanqiang and Zhou, Fenfen and Yin, Ronghui},
  journal={Electronics},
  volume={13},
  number={2},
  pages={243},
  year={2024},
  publisher={MDPI}
}

@article{yang2024multi,
  title={Multi-objective fine-tuning for enhanced program repair with llms},
  author={Yang, Boyang and Tian, Haoye and Ren, Jiadong and Zhang, Hongyu and Klein, Jacques and Bissyand{\'e}, Tegawend{\'e} F and Le Goues, Claire and Jin, Shunfu},
  journal={arXiv preprint arXiv:2404.12636},
  year={2024}
}

%% If your work has an appendix, this is the place to put it.
% \appendix

\end{document}